\documentclass[preprint2]{proto}
\usepackage{times}
\newcommand{\refs}{\par\noindent\hangindent=1pc\hangafter=1}
\usepackage{longtable,lscape}

\newcommand{\kms}{km s$^{-1}$}

\newcommand{\lsun}{\mbox{$L_\odot$}}
\newcommand{\msun}{\mbox{$M_\odot$}}
\newcommand{\AV}{{\it A$_{V}$}}
\newcommand{\Lbol}{$L_{\rm bol}$}

\newcommand{\UNITSOLARLUMI}{\lsun}
\newcommand{\UNITSOLARMASS}{\msun}
\newcommand{\UNITSOLARMASSYEAR}{\mbox{$M_\odot$} yr$^{-1}$}
\newcommand{\UNITVEL}{{\rm km~s$^{-1}$}}

\newcommand{\msunyr}{\mbox{$M_\odot$} yr$^{-1}$}

\newcommand{\Mdot}{{$\dot M$}}

\makeatletter
\newcommand\plotonerot[1]{%
 \typeout{Plotone included the file #1}
 \centering
 \leavevmode
 \includegraphics[angle=90,width={\eps@scaling\columnwidth}]{#1}%
}%
\newcommand\plotonerotb[1]{%
 \typeout{Plotone included the file #1}
 \centering
 \leavevmode
 \includegraphics[angle=270,width={\eps@scaling\columnwidth}]{#1}%
}%
\makeatother

\begin{document}

\title{\textbf{\LARGE Episodic Accretion in Young Stars}}

\author {\textbf{\large Marc Audard}}
\affil{\small\textbf{\textit{University of Geneva}}\vspace*{-2mm}}

\author {\textbf{\large P\'eter \'Abrah\'am}}
\affil{\small\textbf{\textit{Konkoly Observatory}}\vspace*{-2mm}}

\author {\textbf{\large Michael M. Dunham}}
\affil{\small\textbf{\textit{Yale University}}\vspace*{-2mm}}

\author {\textbf{\large Joel D. Green}}
\affil{\small\textbf{\textit{University of Texas at Austin}}\vspace*{-2mm}}

\author {\textbf{\large Nicolas Grosso}}
\affil{\small\textbf{\textit{Observatoire Astronomique de Strasbourg}}\vspace*{-2mm}}

\author {\textbf{\large Kenji Hamaguchi}}
\affil{\small\textbf{\textit{National Aeronautics and Space Administration and University of Maryland, Baltimore County}}\vspace*{-2mm}}

\author {\textbf{\large Joel H. Kastner}}
\affil{\small\textbf{\textit{Rochester Institute of Technology}}\vspace*{-2mm}}

\author {\textbf{\large \'Agnes K\'osp\'al}}
\affil{\small\textbf{\textit{European Space Agency}}\vspace*{-2mm}}

\author {\textbf{\large Giuseppe Lodato}}
\affil{\small\textbf{\textit{Universit\`a Degli Studi di Milano}}\vspace*{-2mm}}

\author {\textbf{\large Marina M. Romanova}}
\affil{\small\textbf{\textit{Cornell University}}\vspace*{-2mm}}

\author {\textbf{\large Stephen L. Skinner}}
\affil{\small\textbf{\textit{University of Colorado at Boulder}}\vspace*{-2mm}}

\author {\textbf{\large Eduard I. Vorobyov}}
\affil{\small\textbf{\textit{University of Vienna and Southern Federal University}}\vspace*{-2mm}}

\author {\textbf{\large Zhaohuan Zhu}}
\affil{\small\textbf{\textit{Princeton University}}}

\begin{abstract}
\baselineskip = 11pt
\leftskip = 0.65in 
\rightskip = 0.65in
\parindent=1pc
{\small In the last twenty years, the topic of episodic accretion has gained significant interest in the star formation community. It is now viewed as a common, though still poorly understood, phenomenon in low-mass star formation. The FU Orionis objects (FUors) are long-studied examples of this phenomenon. FUors are believed to undergo accretion outbursts during which the accretion rate rapidly increases from typically $10^{-7}$ to a few $10^{-4}$~\msunyr, and remains elevated over several decades or more. EXors, a loosely defined class of pre-main sequence stars, exhibit shorter and repetitive outbursts, associated with lower accretion rates. The relationship between the two classes, and their connection to the standard pre-main sequence evolutionary sequence, is an open question: do they represent two distinct classes, are they triggered by the same physical mechanism, and do they occur in the same evolutionary phases? Over the past couple of decades, many theoretical and numerical models have been developed to explain the origin of FUor and EXor outbursts. In parallel, such accretion bursts have been detected at an increasing rate, and as observing techniques improve each individual outburst is studied in increasing detail. 
We summarize key observations of pre-main sequence star outbursts, and review the latest thinking on outburst triggering mechanisms, the propagation of outbursts from star/disk to disk/jet systems, the relation between classical EXors and FUors, and newly discovered outbursting sources -- all of which shed new light on episodic accretion. We finally highlight some of the most promising directions for this field in the near- and long-term. 
 \\~\\~\\~}
\end{abstract}

\section{\textbf{INTRODUCTION}}

Episodic accretion has become a recent focus of attention in the star and planet formation community, turning into a central topic to understand the evolution of protostars and accreting young stars. Initially identified as young stellar objects (YSOs) with strong, long-lived optical outbursts ({\it Herbig}, 1966, 1977), FU Orionis objects (hereafter FUors) have triggered many investigations to understand the eruptive phenomenon. Several reviews have been published  ({\it Herbig}, 1977; {\it Reipurth}, 1990; {\it Hartmann et al.}, 1993; {\it Hartmann}, 1991; {\it Kenyon}, 1995ab; {\it Bell et al.}, 2000;   {\it Hartmann and Kenyon}, 1996;  the specific chapter on the FU Ori phenomenon in {\it Hartmann}, 2008; and the recent review by {\it Reipurth and Aspin}, 2010). The field has  exploded in the last fifteen years thanks to new ground and space facilities.

Observationally FUor candidates --- and their possible short timescale counterparts, EX Lupi objects, dubbed EXors by {\it Herbig}, (1989) ---  have been studied across the electromagnetic spectrum, while theoretical studies have further explored the origin of the outburst mechanism. Erupting young stars are no longer oddities  but are now placed  prominently on the grand scheme of star formation and time evolution of mass accretion rates, from embedded protostars to classical T Tauri stars (CTTS), and eventually weak-line T Tauri stars (WTTS).   In parallel, recent studies have led to doubt as to the need for separate observational classification of FUors and EXors, as discoveries of new outbursting sources have resulted in a less definitive separation. Episodic accretion has also possibly resolved, amongst other issues, the so-called luminosity problem in low-mass embedded protostars.

In this review, we provide a summary of the literature published on the ``historical" FUor and EXor classes and on the theoretical and numerical studies relevant to episodic accretion. We start  from the review by {\it Hartmann and Kenyon}, (1996), although we refer to older studies whenever needed. 
Finally, we emphasize that this review focuses on episodic accretion, i.e., strong  variability due to accretion events: we do not address small-scale variability or variability caused by geometrical effects, clearing of dust, etc., although some of these aspects will be mentioned when observed in parallel with episodic accretion.

\bigskip
\centerline{\textbf{2. OBSERVATIONS}}
\bigskip
\noindent
\textbf{2.1.
Episodic Accretion During Star Formation}
\bigskip

The general picture of the evolution of pre-main sequence accretion ({\it Hartmann and Kenyon}, 1996) suggests that much of the material added to the central star, and the material available for planet-forming disks, is influenced by the frequency and intensity of eruptive bursts followed by long periods of relative quiescence.  It is suspected that this process occurs at {\it all early stages of star formation} after the prestellar core, but becomes observable only as the circumstellar envelope thins.  In one picture, FUors and EXors are part of this continuum.  In this picture, the FUor bursts are longer and stronger compared to the bursts of EXors.  The bursts would occur in repeated cycles and are fueled by additional material falling from the circumstellar envelope to the disk in between bursts, halted by some mechanism, and released in a dramatic flood quasi-periodically.  In an alternative picture, EXors would be a separate phenomenon associated with instabilities in the disks of T Tauri stars (TTS), while FUors span the divide between protostars with disks and envelopes and TTS with disks. Observations reveal a more complicated picture in which strong, long outbursts can  also occur in previously identified CTTS, and EXor-type short outbursts in relatively embedded young stars with envelopes. 

In the next sections, we have kept this  {\it historical, observational} separation between FUors and EXors with the aim to draw commonalities within the classes. We aim to build on the observational and theoretical results to address the validity of the separation and to propose future steps to help determine how and if FUors and EXors are related. 

\bigskip
\noindent
\textbf{2.2.
Characteristics of  FUors}
\bigskip

The initial class of FUors was comprised of  FU Ori, V1057 Cyg, and V1515 Cyg, all showing strong outbursts with amplitudes of several magnitudes, albeit over significantly different timescales and durations ({\it Herbig}, 1977). V1735 Cyg was added to the list shortly thereafter ({\it Elias}, 1978). Although FU Ori has been slowly fading since its 1936 outburst ({\it Kenyon et al.}, 2000), it is still in a high state at present. Notice that V1331 Cyg has often been included among lists of FUors (following {\it Welin,} 1976 who considered it a pre-FUor), but there is no support for such a classification. Since then, several candidates have been added (e.g., {\it Graham and Frogel}, 1985; {\it Brand et al.}, 1986; {\it Eisl{\"o}ffel et al.}, 1990; {\it Staude and  Neckel}, 1991, 1992; {\it Strom and Strom}, 1993; {\it McMuldroch et al.}, 1995; {\it Shevchenko et al.}, 1997; {\it Sandell and Aspin}, 1998; {\it Aspin and Sandell}, 2001; {\it Aspin and Reipurth}, 2003;  {\it Movsessian et al.}, 2003, 2006; {\it Quanz et al.}, 2007ab; {\it Tapia et al.}, 2006; {\it K\'osp\'al et al.}, 2008; {\it Magakian et al.}, 2010, 2013; {\it Reipurth et al.}, 2012). Many objects that are spectroscopically similar to classical FUors but have never been seen to erupt are instead classified as FUor-like objects (in
 analogy with nova-like objects, see {\it Reipurth et al.,} 2002). Many  FUor candidates are significantly more extinguished, with redder/cooler spectral energy distributions (SEDs) than classical FUors, which exhibit relatively evolved/blue SEDs reminiscent of TTS. In Tab.~\ref{tab:fuorexor} we provide a non-exhaustive list of eruptive objects, including FUor-like objects.
 
\begin{deluxetable}{llllllllll}
\tablecolumns{10}
\rotate
\tablewidth{0pc}
\tabletypesize{\tiny}
\tablecaption{Non-exhaustive list of eruptive young stars\label{tab:fuorexor}}
\tablehead{
\colhead{Name}&
\colhead{Type}&
\colhead{Distance}&
\colhead{Onset}&
\colhead{Duration}&
\colhead{\AV}&
\colhead{\Lbol}&
\colhead{\Mdot$_{\rm acc}$}&
\colhead{Companion}&
\colhead{References}\\
\colhead{}&
\colhead{}&
\colhead{(pc)}&
\colhead{(yr)}&
\colhead{(yr)}&
\colhead{(mag)}&
\colhead{(\UNITSOLARLUMI)}&
\colhead{(\msunyr)}&
\colhead{}&
\colhead{}
}
\startdata
RNO 1B&FUor-like&850&\nodata&$>$12&9.2&\nodata&\nodata&Y (RNO 1C, 4")&44,72,78\\
RNO 1C&FUor-like&850&\nodata&\nodata&12.0&\nodata&\nodata&Y (RNO 1B, 4")&44,72\\
V1180 Cas&EXor?&600&2000, 2004&2.5, 7&4.3&0.07 (L)&$>$1.6e-7 (L)&Y? (6.2")&51\\
V512 Per&EXor&300&$>$1988, $<$1990&$>$4&\nodata&66 (L)&\nodata&Y (0.3")&5,12,22\\
PP13S&FUor-like&350&\nodata&\nodata&$\sim$40&30&\nodata&N?&19,73\\
XZ Tau&EXor?&140&1998&$>$3&1.4&0.5&1e-7&Y (0.3")&18,23,31,33\\
UZ Tau E&EXor&140&1921&0.5?&1.5&1.7&1$-$3e-7&Y (SB+4")&23,36,43,56\\
VY Tau&EXor&140&many&0.5$-$2.0&0.85&0.75&\nodata&Y (0.66")&23,36,56\\
LDN 1415 IRS&EXor?&170&$>$2002, $<$2006&\nodata&\nodata&$>$0.13 (L)&\nodata&\nodata&80\\
V582 Aur&FUor&\nodata&$>$1984, $<$1986&$>$26&\nodata&\nodata&\nodata&\nodata&74\\
V1118 Ori&EXor&414&2004,many&$\sim$1.2&0$-$2&1.4 (L), 7$-$25 (H)&2.5e-7 (L), 1e-6 (H)&Y (0.18")&14,36,39,55,59,68\\
Haro 5a IRS & FUor-like & 450 &  \nodata & \nodata & 22 & 50 & \nodata & \nodata & 69  \\
NY Ori&EXor&414&many&$>$0.3&0.3&\nodata&\nodata&N&14,36,39,45,59\\
V1143 Ori&EXor&500&many&$\sim1$&\nodata&\nodata&\nodata&\nodata&39,64\\
V883 Ori&FUor-like&460&\nodata&\nodata&\nodata&400&\nodata&\nodata&72,81\\
Reipurth 50 N IRS 1&FUor-like&460&\nodata&\nodata&\nodata&300&\nodata&\nodata&16,81\\  
V2775 Ori&FUor-like&420&$>$2005, $<$2007&$>$ 5&8$-$12&2-4.5 (L), 22-28 (H)&2e-6 (L), 1e-5 (H)&Y? (11")&24\\
FU Ori&FUor&450&1936&$\sim$100&1.5$-$2.6&340$-$500&\nodata&Y (0.5")&1,34,77\\
V1647 Ori&EXor?&400&1966,2003,2008&0.4$-$1.7,2.5,$>$4.3&8$-$19&3.5$-$5.6,34$-$44&6e-7,4e-6$-$1e-5&\nodata&3,4,7,8,9,10,15,25,63,83,84\\
AR 6A & FUor-like &800 &  \nodata & $>$13& 18 & 450 & \nodata & Y (AR 6B, 2.8") & 13\\
AR 6B & FUor-like & 800 & \nodata & \nodata & $>$18 & \nodata &  \nodata & Y (AR 6A, 2.8") & 13\\
V900 Mon&FUor-like&1100&$>$1953, $<$2010&$>$16&13&106 (H)&\nodata&N&70\\
Z CMa&FUor&930$-$1100&many&5$-$10&1.8$-$3.5&400$-$600&1e-3&Y (0.1")&34,35,46,53,72,85\\
BBW 76&FUor-like&1700&$<$1900&$\sim$40&2.2&287&7.2e-5&N&1,27,34\\
V723 Car&EXor&\nodata&\nodata&\nodata&\nodata&\nodata&\nodata&\nodata&82\\
GM Cha&EXor?&160&many&$>$1.9&$\gtrsim$13&$>$1.5&1e-7&Y (10")&66\\
EX Lup&EXor&155&2008,many&$<$1&0&0.7,2&4e-10,2e-7&Y? (BD)&2,6,30,36,37,38,54,76\\
V346 Nor&FUor&700&$\sim$1980&$>$5&$>$12&135&\nodata&N&1,26,27,34,67\\
OO Ser&FUor-like&311&1995&$>$16&42&4.5 (L), 26-36 (H)&\nodata&N&32,40,41,42,48\\
Parsamian 21&FUor-like&400&\nodata&\nodata&8?&3.4, 10&\nodata&N?&20,47,79\\
V1515 Cyg&FUor&1000, 1050&$\sim$1950&$\sim$30&2.8$-$3.2&200&3.5e-5&\nodata&1,27,34,77\\
PV Cep&EXor?&325&repetitive&$\sim$ 2&12.0&41 (L), 100 (H)&2e-7$-$2.6e-6 (L), 5.2e-6 (H)&\nodata&52\\
V2492 Cyg&EXor?&600&$>$2009, $<$2010&$>$3&6$-$12, 10$-$20&14 (L), 43 (H)&2.5e-7 (H)&\nodata&21,49,50\\
HBC 722&FUor&600&2009&$>$4&3.4, 3.1&0.7$-$0.85 (L), 8.7$-$12 (H)&1e-6 (H)&\nodata&29,49,60\\		
V2494 Cyg&FUor-like&700$-$800&$>$1952, $<$1989&$>$20&5.8&14$-$18&\nodata&\nodata&57,69\\		
V1057 Cyg&FUor&600, 700&1970&$\sim$10&3.0$-$4.2&250$-$800&4.5e-5&N&1,27,34,77\\
V2495 Cyg&FUor&800&1999&$>$8&\nodata&\nodata&\nodata&\nodata&11,62\\
RNO 127&FUor&800&1999&$>$6&\nodata&\nodata&\nodata&\nodata&61,62\\	
V1735 Cyg&FUor&900&$>$1957, $<$1965&$>$20&8.0$-$10.8&235&\nodata&Y? (20-24")&1,34,77\\
HH 354 IRS & FUor-like & 750 &  \nodata & \nodata & \nodata & 73.0 & \nodata & \nodata & 20,69\\
V733 Cep&FUor-like&800&$>$1953, $<$1984&$>$38&8&135 (H)&\nodata&\nodata&65,71\\
\enddata
\tablecomments{
~L: quiescent or low state, H: outburst or high state, Companion: The parentheses show the angular separation of the companion from FUor/EXor. References: 1: {\it {\'A}brah{\'a}m et al.,} (2004b); 2: {\it {\'A}brah{\'a}m et al.,} (2009); 3: {\it {\'A}brah{\'a}m et al.,} (2004a); 4: {\it Andrews et al.,} (2004); 5: {\it Anglada et al.,} (2004); 6: {\it Aspin et al.,} (2010); 7: {\it Aspin et al.,} (2009b); 8: {\it Aspin et al.,} (2006); 9: {\it Aspin,} (2011b); 10: {\it Aspin et al.,} (2008); 11: {\it Aspin et al.,} (2009c); 12: {\it Aspin and Sandell,} (1994); 13: {\it Aspin and Reipurth,} (2003); 14: {\it Audard et al.,} (2010); 15: {\it {Brice{\~n}o} et al.,} (2004); 16: {\it Casali,} (1991); 17: {\it Chavarria,} (1981); 18: {\it Coffey et al.,} (2004); 19: {\it Cohen et al.,} (1983); 20: {\it Connelley and Greene,} (2010); 21: {\it Covey et al.,} (2011); 22: {\it Eisl\"offel et al.,} (1991); 23: {\it Elias,} (1978); 24: {\it Fischer et al.,} (2012); 25: {\it Gibb,} (2008); 26: {\it Graham and Frogel,} (1985); 27: {\it Green et al.,} (2006); 28: {\it Green et al.,} (2013b); 29: {\it Green et al.,} (2011); 30: {\it Grosso et al.,} (2010); 31: {\it Haas et al.,} (1990); 32: {\it Haisch et al.,} (2004); 33: {\it Hartigan and Kenyon,} (2003); 34: {\it Hartmann and Kenyon,} (1996); 35: {\it Hartmann et al.,} (1989); 36: {\it Herbig,} (1989); 37: {\it Herbig,} (1977); 38: {\it Herbig,} (2007); 39: {\it Herbig,} (2008); 40: {\it Hodapp,} (1996); 41: {\it Hodapp,} (1999); 42: {\it Hodapp et al.,} (2012); 43: {\it Jensen et al.} (2007); 44: {\it Kenyon et al.,} (1993); 45: {\it K\"ohler et al.,} (2006); 46: {\it Koresko et al.,} (1991); 47: {\it K{\'o}sp{\'a}l et al.,} (2008); 48: {\it K{\'o}sp{\'a}l et al.,} (2007); 49: {\it K{\'o}sp{\'a}l et al.,} (2011a); 50: {\it K{\'o}sp{\'a}l et al.,} (2013); 51: {\it Kun et al.,} (2011a); 52: {\it Kun et al.,} (2011b); 53: {\it Leinert and Haas,} (1987); 54: {\it Lombardi et al.,} (2008); 55: {\it Lorenzetti et al.,} (2006); 56: {\it Lorenzetti et al.,} (2007); 57: {\it Magakian et al.,} (2013); 58: {\it McMuldroch et al.,} (1993); 59: {\it Menten et al.,} (2007); 60: {\it Miller et al.,} (2011); 61: {\it Movsessian et al.,} (2003); 62: {\it Movsessian et al.,} (2006); 63: {\it Muzerolle et al.,} (2005); 64: {\it Parsamian et al.,} (1987); 65: {\it Peneva et al.,} (2010); 66: {\it Persi et al.,} (2007); 67: {\it Prusti et al.,} (1993); 68: {\it Reipurth et al.,} (2007b); 69: {\it Reipurth and Aspin,} (1997); 70: {\it Reipurth et al.,} (2012); 71: {\it Reipurth et al.,} (2007a); 72: {\it Sandell and Weintraub,} (2001); 73: {\it Sandell and Aspin,} (1998); 74: {\it Semkov et al.,} (2011); 75: {\it Shevchenko et al.,} (1991); 76: {\it Sipos et al.,} (2009); 77: {\it Skinner et al.,} (2009); 78: {\it Staude and Neckel,} (1991); 79: {\it Staude and Neckel,} (1992); 80: {\it Stecklum et al.,} (2007); 81: {\it Strom et al.,} (1993); 82: {\it Tapia et al.,} (2006); 83: {\it Teets et al.,} (2011); 84: {\it Venkata Raman et al.,} (2013);  85: {\it Whelan et al.,} (2010)}
\end{deluxetable}

In the optical, classical FUors  exhibit similar spectra, with F/G supergiant spectral types but broad lines compared to TTS, and little indication of magnetospheric accretion.  In contrast, they show K/M supergiant spectral types in the near-infrared. A supergiant spectrum can also be seen in the ultraviolet ({\it Kravtsova et al.}, 2007).
They are associated with reflection nebulae and are associated with a moderate level of extinction (A$_V$ $\sim$ 1.8 -- 3.5 mag).  In the near-infrared, FUors show first-overtone CO absorption at 2.2 $\mu$m.  The Fe~\textsc{i}, Li~\textsc{i}, and Ca~\textsc{i} optical lines, as well as the near-infrared CO lines, are typically double-peaked and show line broadening that is kinematically consistent with a rotating disk origin ({\it Hartmann and Kenyon}, 1996), although a wind may also be required to explain these profiles ({\it Eisner and Hillenbrand}, 2011). A different origin for the optical line splitting was put forward by {\it Petrov and Herbig}, (1992) and {\it Petrov et al.}, (1998) who argued that the profile can be explained simply by the presence of central emission cores in the broad absorption lines (see also {\it Herbig et al.}, 2003). Similarly, a model invoking a large dark polar spot was put forward to explain the line profiles in the optical for FU Ori ({\it Petrov and Herbig}, 2008). Despite the controversy at optical wavelengths (which may in fact relate to different aspects of the same underlying phenomenon, see {\it Kravtsova et al.}, 2007), the accretion disk model  generally works well  (e.g., {\it Hartmann et al.}, 2004).

In the infrared, the similarity between classical FUors begins to waver, as FU Ori exhibits pristine silicate emission features and a relatively blue SED consistent with that of a mildly flared disk ({\it Green et al.}, 2006; {\it Quanz et al.}, 2007c), while V1515 Cyg and V1057 Cyg are closer to flat-spectrum sources with weaker silicate emission.   They are further distinguished in the far-infrared/submillimeter where envelopes often dominate SEDs: FU Ori and V1515 Cyg have  weak continuum beyond 100 $\mu$m and very little CO line emission, while V1057 Cyg shows warm CO gas and substantially stronger continuum ({\it Green et al.}, 2013b). Nevertheless, near-infrared interferometry shows that classical FUors all show significant contributions from envelopes  ({\it Millan-Gabet et al.}, 2006). Such envelopes have masses of a few tenths of a solar mass and are comparable to Class I protostars rather than Class II  stars ({\it Sandell and Weintraub}, 2001; {\it P\'erez et al.}, 2010). 

During their outbursts, classical FUors have bolometric luminosities of $100-300~\lsun$, with accretion rates between 10$^{-6}$ and 10$^{-4}$~\msunyr\  (Sect. 4.4).  One of the challenges is identifying candidate FUors before they occur: V1057 Cyg is one of only two identified FUors (with HBC 722, discussed below) with a pre-outburst optical spectrum ({\it Herbig}, 1977). It was found  to have optical emission lines typical of a TTS.  Modern  large spectral surveys of active star-forming regions should increase the likelihood of serendipitous pre-outburst observations (Sect. 5).

Since  few  FUor outbursts have been detected, the question arises as to whether some existing YSOs may have previously undergone undetected FUor eruptions.  For instance some driving sources of Herbig-Haro (HH) objects might be  FUor-like, because their near-infrared spectra are substantially similar to those of FUors ({\it Reipurth and Aspin}, 1997;  {\it Greene et al.}, 2008).
 These authors noted that  young stars with FUor-like characteristics might be more common than projected from the relatively few known classical FUors.

\bigskip
\noindent
\textbf{2.3.
FUor Temporal Behavior}
\bigskip

\textit{2.3.1. Long-term evolution.}
In the optical, the initial brightening of $\sim 5$~mag is followed by a longer plateau phase, during which classical FUors display a relatively long decay timescale: FU Ori has faded by about 1 mag since its 1936 outburst, at a rate of 14 mmag yr$^{-1}$ ({\it Kenyon et al.}, 2000). BBW 76 (also known as Bran 76) showed a decay of about 23 mmag yr$^{-1}$ in $V$ between 1983 and 1994, with a slow-down in the infrared around 1989, although a historical search indicated that BBW 76 has remained at a similar brightness level since at least 1900  ({\it Reipurth et al.}, 2002). 
To document the long-term flux evolution of FUors, {\it \'Abrah\'am et al.}, (2004b) compared 1--100 $\mu$m photometric observations, obtained by {\it ISO} in 1995-98, with earlier data taken from the {\it IRAS} catalog (from 1983). Both satellite data sets were supplemented by contemporaneous ground-based infrared observations from the literature. In two cases (Z CMa, Parsamian 21) no significant difference between the two epochs was seen. V1057 Cyg, V1515 Cyg and V1735 Cyg became fainter at near-infrared wavelengths while V346 Nor had become slightly brighter. At $\lambda \geq$ 60 $\mu$m most of
the sources remained constant; only V346 Nor seemed to fade. A detailed case study of V1057 Cyg revealed that the long-term evolution of the system at near- and mid-infrared wavelengths was consistent with model predictions of {\it Kenyon and Hartmann}, (1991) and {\it Turner et al.}, (1997), except at wavelengths longward of 25~$\mu$m.

\textit{2.3.2. Variability and episodic accretion.}
In addition to broad lightcurve evolution, FUors exhibit a variety of short-timescale (hours to days) variability, both semi-periodic and stochastic.  
{\it Kenyon et al.}, (2000) present rapid cadence photometry  of classical FUors.  They find evidence for low amplitude ``flickering'' ($\sim$ 0.03-0.2 mag) over a time scale of $\leq 1$~d which they attribute to the inner accretion disk.  Similar results were found by {\it Clarke et al.}, (2005). From high spectral resolution optical observations of FU Ori, {\it Powell et al.}, (2012) confirmed the modulation of  wind lines ($P\sim$14 days) and photospheric lines ($P\sim$3.5 days) discovered by {\it Herbig et al.}, (2003), while  {\it Siwak et al.}, (2013) used {\it MOST} photometry   and found the possible existence of 2--9~d quasi-periodic features in FU Ori, which they interpreted as plasma condensations or localized disk heterogeneities. 
The flickering and presence of periodicity  was also found in other erupting stars, such as V1647 Ori ({\it Bastien et al.}, 2011).  {\it Green et al.}, (2013a)  found a rotational period for HBC 722 of 5.8 days, and a variety of sub-periods between 0.9 and 1.3 days, with a peak in the periodogram at 1.28 days.  This shorter period was attributed to instability near the inner disk edge (see also Sect. 3.4.3).

In any case, episodic accretion can be distinguished from ``normal'' TTS variability that results from either small-scale accretion events, geometrical effects, or magnetic activity ({\it Costigan et al.}, 2012; {\it Chou et al.}, 2013; {\it Scholz et al.}, 2013). Indeed, accreting TTS tend to vary in mass accretion rates by 0.37-0.83~dex or less, depending on the accretion diagnostic. Such variability occurs over time scales of $\leq 15$ months, with a major part of variability occurring over shorter time scales, 8--25 days, i.e., comparable to stellar rotational periods. In contrast, episodic accretion is characterized by stronger variations in mass accretion rate and occurs on longer time scales (i.e., months to years).


\bigskip
\noindent
\textbf{2.4.
Cold Environments of FUors}
\bigskip

(Sub-)millimeter observations   originally suggested that classical FUors are extremely young, more similar to Class\,I protostars than to Class\,II stars ({\it Sandell and Weintraub}, 2001). These single-dish continuum maps showed resolved emission and typically elongated disk-like morphology, with sizes
of several thousand AU (e.g., {\it Dent et al.}, 1998; {\it Henning et al.}, 1998).  The mass in these structures was estimated at a few tenths of \msun. Such reservoirs are sufficient to
replenish circumstellar disks after episodes of rapid accretion. Molecular outflows were generally detected, except for the most evolved FUors ({\it Evans et al.}, 1994; {\it Moriarty-Schieven et al.}, 2008), with mass loss rates between a few times 10$^{-8}$ and 10$^{-6}$~\msunyr, without correcting for optical depth; mass loss rates likely peak at 10$^{-5}$~\msunyr. 
Dense gas traced by HCO$^+$ was also detected ({\it McMuldroch et al.}, 1993, 1995; {\it Evans et al.}, 1994).  Newly obtained line profiles obtained by {\it Green et al.}, (2013b) with {\it Herschel} are consistent with the above results. However, with single-dish observations, it is often difficult to determine whether the progenitor of the outflow is the FUor or in fact nearby young protostars, such as in the case of HBC 722, which remained undetected in the continuum, with an upper limit of 0.02 \msun\ for the mass of the disk ({\it Dunham et al.}, 2012), low enough to rule out gravitational instabilities in the disk as the driving mechanism for  the outburst.

Further evidence of gas emission associated with FUors was obtained with {\it ISO} by {\it Lorenzetti et al.}, (2000) who found [OI], [CII], and [NII], and some cold molecular lines, both on and off-source. All FUors observed with {\it Herschel} by {\it Green et al.}, (2013b) exhibited strong [O I] 63 $\mu$m emission, indicative of high mass loss rates or UV excitation. On the other hand, the warm CO emission lines, generally observed in Class 0/I protostars, were not detected,  with only V1057 Cyg showing compact high-$J$ CO emission.

Interferometric observations are, however, required to distinguish between the circumstellar envelope and cloud emission.  {\it K\'osp\'al}, (2011) detected strong, narrow $^{13}$CO J=1--0  line emission in Cygnus FUors. The emission was spatially resolved in all cases, with deconvolved sizes of a few thousand AU. For V1057\,Cyg, the emitting area was rather compact, suggesting that the origin of the emission is a circumstellar envelope surrounding the central star. For V1735\,Cyg, the $^{13}$CO emission was offset from the stellar position, indicating that the source of this emission might be a small foreground cloud, also responsible for the high reddening of the central star. The $^{13}$CO emission towards
V1515\,Cyg was the most extended in the sample, and it apparently coincided with the ring-like optical reflection nebula associated with V1515\,Cyg.

Future interferometric observations in the (sub-)millimeter will better disentangle the envelope and disk emissions, and  constrain the mass reservoir around FUors.

\bigskip
\noindent
\textbf{2.5.
Observational Classification Schema for FUors}
\bigskip

There are a number of methods used to classify young stars, such as the infrared spectral index $\alpha$ ({\it Andr\'e et al.}, 1993; {\it Greene et al.}, 1994) or the bolometric temperature $T_{\rm bol}$ ({\it Chen et al.}, 1995; {\it Robitaille et al.}, 2006).  Some FUors exhibit Class II SEDs, while others exhibit  flat SEDs (characterized as having 350 K $<$ $T_{\rm bol}$ $<$ 950 K, {\it Evans et al.}, 2009; {\it Fischer et al.}, 2013), near the Class I/II boundary ({\it Green et al.}, 2013b).  The envelope mass derived from (sub-)millimeter dust continuum can also be used to discriminate between evolutionary Stages. A thorough discussion on the distinction between Stages and the observationally-defined infrared Class is provided in the chapter on protostellar evolution by {\it Dunham et al.}

\begin{figure*}[!th]
 \epsscale{2}
 \plottwo{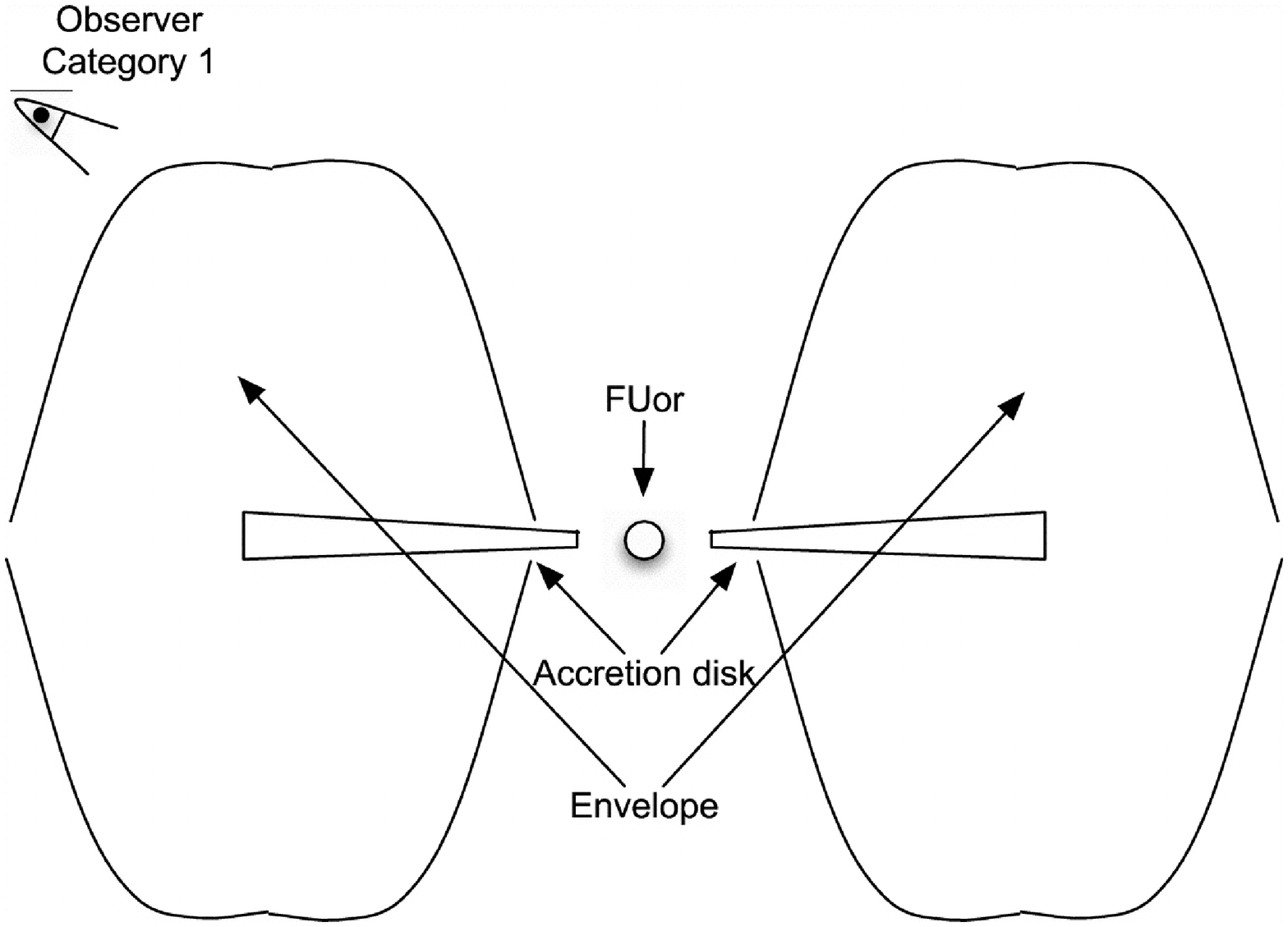}{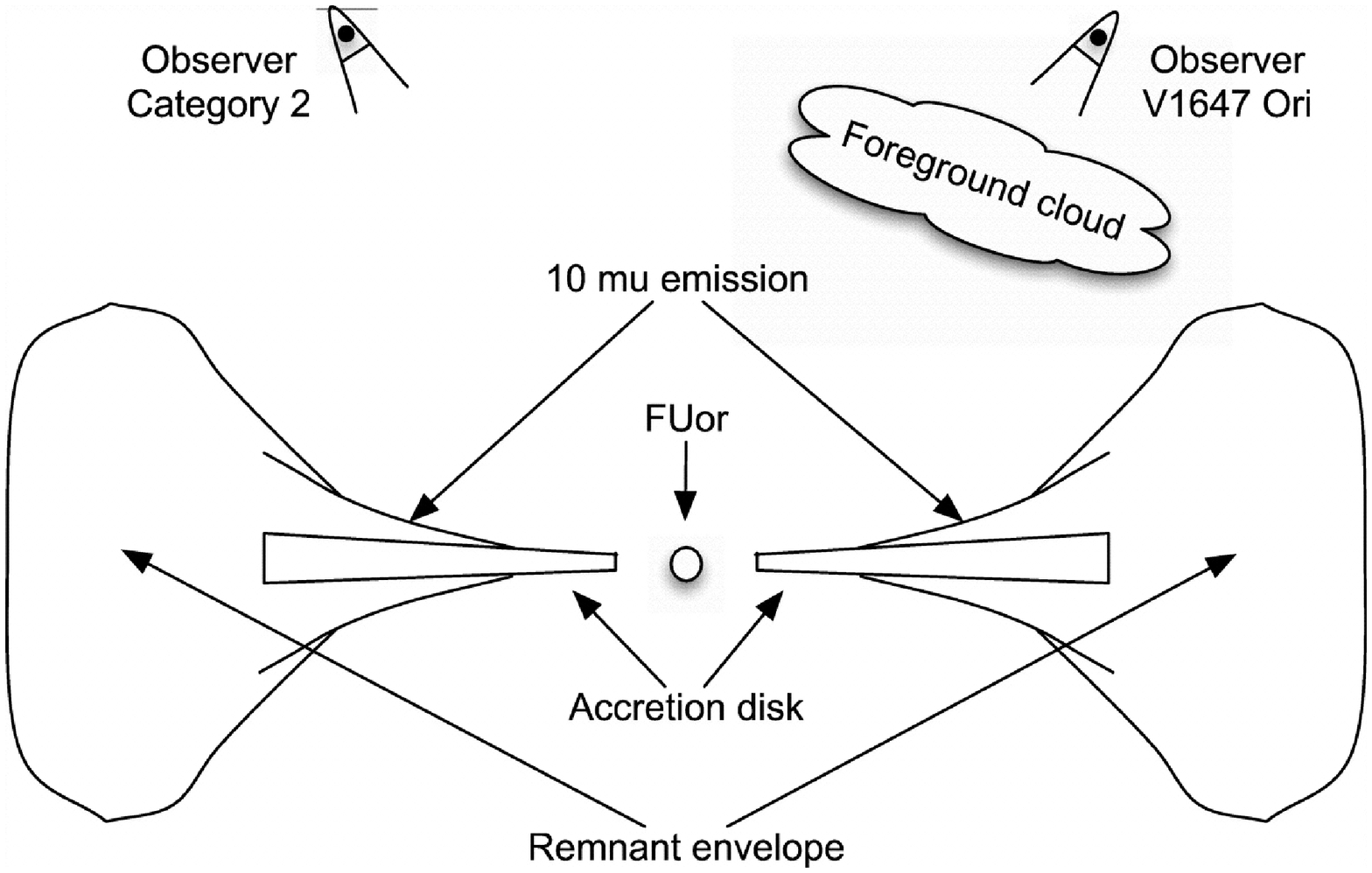}
 \caption{\small Sketch of the two categories of FUors ({\it Quanz et al.}, 2007c). The first category is related to embedded sources with silicate features in absorption and younger ages than FUors of the second category, which show silicate features in emission. Reproduced by permission of the AAS.}\label{fig:quanz07categ}
 \end{figure*}

Several studies have focused on the silicate feature around 10~$\mu$m  and  on ice properties ({\it Polomski et al.,} 2005; {\it Sch\"utz et al.}, 2005; {\it Green et al.}, 2006; {\it Quanz et al.}, 2007c). The silicate feature can be found either in absorption   or in emission. The FUors with silicate absorption generally show an amorphous silicate structure, similar to the interstellar medium, although some extra emission or large silicates can be found. In addition, H$_2$O, methanol, and CO$_2$ ice absorption is detected. In contrast,  FUors with silicate in emission can show evidence for absorption in the disk photosphere from blended H$_2$O vapor absorption (5-8 $\mu$m; {\it Green et al.}, 2006). The silicate feature in emission does not show evidence of crystals, with some FUors showing weak, pristine silicate features (e.g., V1515 Cyg and V1057 Cyg). This led {\it Quanz et al.}, (2007c) to propose that FUors can be classified in two categories, defined by silicate absorption vs.~emission.  The emission features arise from the disk surface layers, and represent evidence for grain growth but no processing (Fig.~\ref{fig:quanz07categ}).  Objects with silicate in absorption are likely young FUors, still embedded in an envelope, whereas objects with silicates in emission are likely  more evolved FUors with (partially) depleted envelopes.

{\it Green et al.}, (2013b) found that the mid-infrared sequence proposed by {\it Quanz et al.}, (2007c) broadly described the continuum of FUors for which there was sufficient \mbox{(sub-)}millimeter data.  However, the line observations provide a different picture from the continuum, as FUors may not be well characterized by the Stage I/II sequence.  While FU Ori and V1515 Cyg, two of the most evolved FUors, were found to have little warm ($\sim$ 100 K) gas, V1057 Cyg exhibited CO rotational lines typical of Class 0/I protostars.  If this CO originates in the shocked layer of the envelope surrounding the outflow, FU Ori's lack of emission can be explained by a tenuous envelope; however the mid-infrared similarity between V1515 Cyg and V1057 Cyg (in contrast with FU Ori) would not predict the difference in their submillimeter emission lines.  The foregoing makes apparent the difficulty of classifying FUors in the Stage I/II sequence.


\bigskip
\noindent
\textbf{2.6.
Characteristics of EXors}
\bigskip

Classical EXors ({\it Herbig}, 1989, 2008) were initially defined as a list of variable stars showing large-scale outbursts and TTS spectra. The prototype, EX Lup, largely influenced the definition of the class, with its repetitive, short-lived outbursts, M-type dwarf spectrum in quiescence, and absence of features typical of FUors (see Sect.~2.2). The original list of EXors ({\it Herbig}, 1989) has changed little since that time. New potential EXors were found, although their classifications as EXor or FUor are sometimes unclear (e.g., {\it Eisl{\"o}ffel et al.}, 1991; {\it Aspin and Sandell}, 1994;  {\it Stecklum et al.}, 2007;  {\it Persi et al.}, 2007; {\it Kun et al.}, 2011a).  Abundant photometric observations during and after outbursts
indicated brightness increases of several magnitudes over  durations of several months (e.g.,  {\it Coffey et al.}, 2004; {\it Audard  et al.}, 2010). Bolometric luminosities are about 1--2~\lsun\  in quiescence (including about 0.3--0.5~\lsun\   of stellar photospheric luminosity) and peak around several \lsun\  to a few tens of \lsun   ({\it Lorenzetti et al.}, 2006; {\it Audard et al.}, 2010; {\it Sipos et al.}, 2009; {\it Aspin et al.}, 2010). Coverage tends to become less frequent after outburst, preventing precise determinations of EXor outburst durations. Nevertheless, eruptions occur frequently in the same objects, with separations of a few years between outbursts, and durations of about 1--2 years (e.g., {\it Herbig}, 2008). 
Oscillations ($\approx$30--60~days), longer than the rotation period,  can be detected during outbursts and can be explained by magnetospheric accretion models of trapped disks near corotation (e.g., {\it Acosta-Pulido et al.}, 2007; {\it D'Angelo and Spruit}, 2012).

EXors show absorption spectra typical of K or M dwarfs with T Tauri-like emission spectral features during their minimum light ({\it Parsamian and Mujica}, 2004; {\it Herbig}, 2008).  Optical and near-infrared SEDs during outbursts can be well fitted with an extra thermal component, e.g., a single blackbody component with temperatures varying from 1000 K to 4500 K  and emitting radii of 0.01 to 0.1 AU, typical of inner disk regions ({\it Lorenzetti et al.,} 2012). Alternatively, the outburst SED can be fitted with a hotspot model whose emission dominates in the optical ({\it Audard et al.}, 2010; {\it Lorenzetti et al.,} 2012). The coverage factor can be large during the outburst, although {\it Lorenzetti et al.,} (2012) argue that it does not exceed about 10~\% of the stellar surface. In any case, color variations observed during outbursts cannot be explained  by extinction effects alone. Signatures of infall and outflow are detected in Na I D$_{1,2}$, in a similar fashion as in CTTS. Near-infrared spectra show line emission from hydrogen recombination lines, with CO overtone features and weaker atomic features commonly detected both in emission and absorption and variable on short timescales ({\it Lorenzetti et al.}, 2009). 

The SEDs of EXors span the Class I/II divide symmetrically, with a peak in the ``flat spectrum'' category ({\it Giannini et al.}, 2009). In the mid-infrared, EXors show the 10$\,\mu$m silicate feature in emission ({\it Audard et al.}, 2010; {\it K\'osp\'al et al.}, 2012). Some of them show wavelength-independent flux changes, probably due to varying accretion. Others are more variable in the 10$\,\mu$m silicate feature than in the adjacent continuum,
which can be interpreted as possible structural changes in the inner disk. 

\bigskip
\noindent
\textbf{2.7. New Eruptive Stars}
\bigskip

\begin{figure}[!t]
 \epsscale{1.}
  \plotone{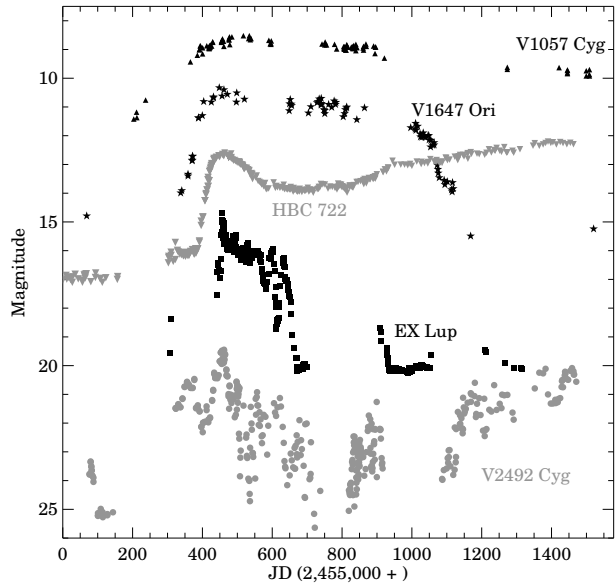}
 \caption{\small  Comparison light curves for the FUor V1057 Cyg, the intermediate case V1647 Ori, the new sources HBC 722 and VSX J2025126.1+440523 (also known as PTF 10nvg or V2492 Cyg), and the classical EXor EX Lup, showing the continuum of outburst durations. Adapted from {\it K\'osp\'al et al., } (2011a).\label{fig:ilight}
 }  
 \end{figure}

Several newly discovered eruptive young stars have been found, many in the last decade. Some of them often show spectral characteristics typical of FUors, but smaller luminosities (see Tab.~\ref{tab:fuorexor}).  HBC 722 (also known as V2493 Cyg) started with a  $0.6-0.85~\lsun$ luminosity and a SED typical of CTTS, then its luminosity rose to $4-12~\lsun$ in outburst ({\it Semkov et al.}, 2010; {\it Miller et al.}, 2011; {\it K\'osp\'al et al.}, 2011a). Similarly, V2775 Ori changed its bolometric luminosity from $4.5~\lsun$ to $\approx 51~\lsun$ ({\it Caratti o   Garatti et al.}, 2011; {\it Fischer et al.}, 2012).  LDN 1415 IRS could also be a low-luminosity erupting star ($0.13~\lsun$ in low state; {\it Stecklum et al.}, 2007), although its status as FUor or EXor is yet unclear.  Moderate luminosities are also observed in OO Ser ({\it Hodapp et al.}, 1996, 2012), with $26-36~\lsun$ in outburst, whereas V733 Cep displayed stronger luminosity ($135~\lsun$) and has  slowly faded since 2007 ({\it Peneva et al.}, 2010). The luminosity of V2494 Cyg (=HH 381 IRS) is difficult to constrain ($14-45$~\lsun) in light of the uncertainty in its distance ({\it Magakian et al.}, 2013). The recently discovered V1647 Ori also showed moderate luminosity at peak (about $20-40~\lsun$; {\it Andrews et al.}, 2004; {\it Muzerolle et al.}, 2005; {\it Aspin}, 2011b), although it may more closely resemble an EXor than a FUor. Some new sources are young (typically Class I) and deeply embedded, hidden behind thick envelopes  (e.g., OO Ser: {\it K\'osp\'al et al.}, 2007; V900 Mon:  {\it Reipurth at al.}, 2012). Others have no detectable envelopes, suggesting a relatively evolved state (e.g., V733 Cep: {\it Reipurth et al.}, 2007a; HBC 722: {\it Green et al.}, 2011, 2013c; {\it Dunham et al.}, 2012), despite sometimes displaying  heavy extinction due to surrounding clouds. 
 
Even more interestingly, time scales for outbursts were found {\it in between} those of classical EXors and FUors (Fig.~\ref{fig:ilight}). OO Ser was in outburst for about 8 yrs ({\it Hodapp et al.}, 1996, 2012). Together with its moderate luminosity, it bears resemblance to V1647 Ori, which has been twice in outburst since 2004 (see Sect.~2.9.3). HBC 722 also displayed quite peculiar behavior: it started fading after peak brightness with a rate much faster than typical for FUors, but the fading stopped, and the object started brightening again, with no clear signs that its outburst will end any time soon ({\it Semkov et al.}, 2012; Fig.~\ref{fig:ilight}).  These discoveries demonstrate that eruptive phenomena span a considerable range in evolutionary state, envelope mass, stellar mass, time behavior, and accretion rates. 

\bigskip
\noindent
\textbf{2.8.
Re-arrangements in the Circumstellar Matter}
\bigskip

Several unusual eruptive stars were recently identified: sources where the brightening may be due to the combination of two effects ({\it Hillenbrand et al.}, 2013). One effect is an intrinsic brightening related to enhanced accretion, as in all eruptive stars (e.g., {\it Sicilia-Aguilar et al.}, 2008). The other effect is a
dust-clearing event that reduces the extinction along the line of sight, such as in UX Ori-type stars (e.g., {\it Xiao
  et al.}, 2010; {\it Chen et al.}, 2012; {\it Semkov and Peneva}, 2012).  Tab.~1 does not include sources for which extinction effects dominate the time variability (e.g., GM Cep, V1184 Tau). Accretion and extinction changes often happen synchronously, suggesting that they are physically linked. The extinction can reach high values (e.g., {\it Covey et al.}, 2011; {\it  K\'osp\'al et al.}, 2011a). The variations are likely due to obscuring structures lying close to the stars (i.e., within a few tenths of an AU, {\it K\'osp\'al et al.}, 2013). Interestingly, the objects in question can also be highly embedded Class I objects, such as V2492 Cyg, and they can display rich and variable emission-line spectra like EXors ({\it Aspin}, 2011a). The intermediate-mass star PV Cep, originally classified as an EXor by {\it Herbig}, (1989) on the basis of its large outburst in 1977--1979 ({\it Cohen et al.}, 1981), despite its mass, embeddedness and higher accretion rate than typical EXors, also shows variability resulting from an interplay between variable accretion and circumstellar extinction, hinting at a rapidly changing dust distribution close to the star ({\it Kun et al.}, 2011b; {\it   Lorenzetti et al.}, 2011; see also {\it Caratti o Garatti et al.}, 2013). 

These new observations indicate that structural changes often happen in the innermost part of the disk or circumstellar matter, typically within 1\,AU, supporting the conclusion that these changes are related to the outburst. It is remarkable that all of the objects described above show EXor-like, repetitive outbursts. This implies that whatever structural change happens in the disk, it should be reversible on a relatively short timescale, setting strong constraints on the  physical processes involved.

\bigskip
\noindent
\textbf{2.9.
Multi-Wavelength Studies of Outbursts}
\bigskip

Recent outbursts in classical EXors  and in new eruptive young stars triggered strong interest in obtaining detailed follow-up observations at all possible wavelengths. We focus  on three well-studied objects --- the classical EXors EX Lup and V1118 Ori, and the recently discovered V1647 Ori --- that reflect the diversity of outburst properties.


\begin{figure*}[!t]
 \epsscale{1.9}
 \plotonerotb{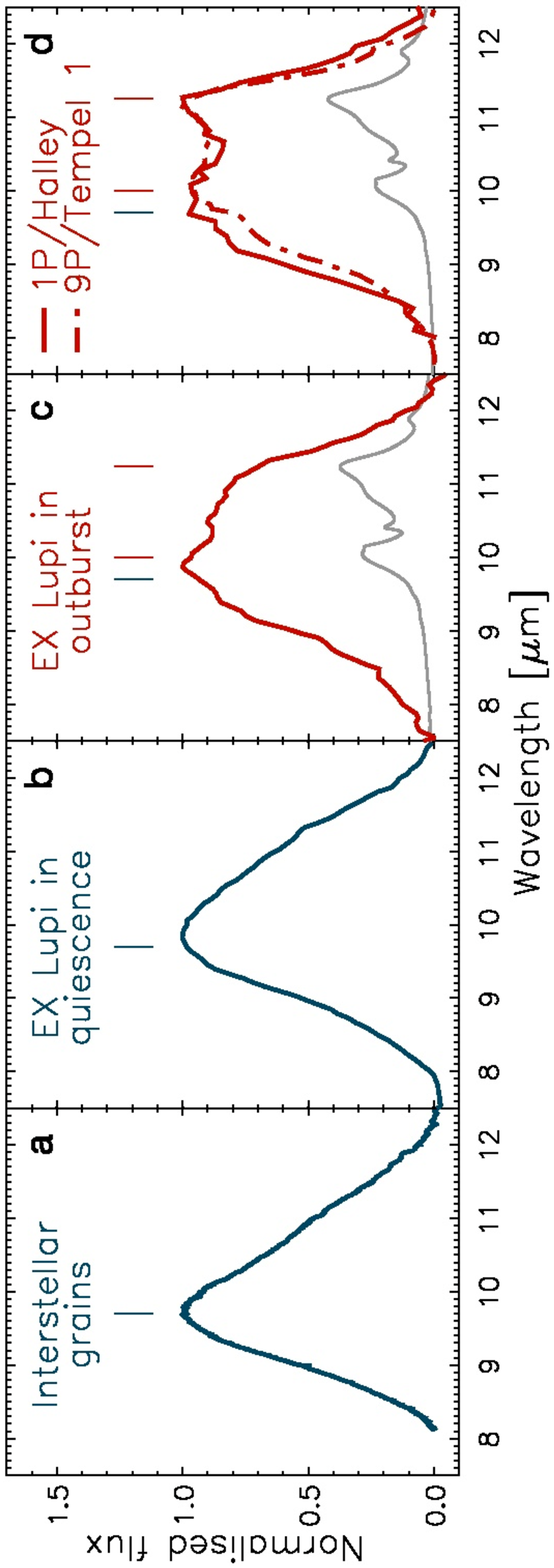}
 \caption{\small Silicate emission spectra for the interstellar medium (a), EX Lup in quiescence (b), during its 2008 outburst (c), and for two comets (d). The bottom curves in panels (c) and (d) show the emissivity curve of pure forsterite, for grain temperatures of 1250 K and 300 K, respectively. The outburst spectrum of EX Lup shows evidence of forsterite, not observed during quiescence, and produced through thermal annealing in the surface layer of the inner disk by heat from the outburst. Adapted from  {\it \'Abrah\'am et al.,} (2009).\label{fig:exlupsil}
}  
 \end{figure*}
\textit{2.9.1. The prototype EX\,Lupi.}
EX Lup is a young low-mass ($0.6~\msun$) M0V star ({\it Gras-Vel\'azquez and Ray}, 2005), with a quiescent $L_{\rm bol} = 0.7~\lsun$. Its disk has a mass of  0.025\,M$_{\odot}$, with an inner hole within 0.2--0.3~AU, and an outer radius of 150\,AU ({\it Sipos et al.}, 2009). There is no hint of any envelope. This prototype of the EXor class shows small amplitude variations in quiescence, but has displayed several  outbursts ({\it Lehmann et al.}, 1995;  {\it Herbig et al.}, 2001; {\it Herbig}, 2007) during which the photospheric spectrum is veiled by a hot continuum, the equivalent widths of the optical emission lines decrease, inverse P\,Cygni absorption components appear at the higher Balmer lines, the emission-line structures undergo striking variations, and many emission lines exhibit both a narrow and a broad line profile component. All these signatures indicate mass infall in a magnetospheric accretion event. In quiescence, permitted emission lines and numerous metallic lines are detected ({\it K\'osp\'al et al.}, 2011b; {\it Sicilia-Aguilar et al.}, 2012), likely originating from a hot (6500~K), dense, non-axisymmetric, and non-uniform accretion column that displays velocity variations along the line of sight on timescales of days. Further evidence supporting a magnetospheric accretion model is given by the spectro-astrometry of near-infrared hydrogen lines ({\it K\'osp\'al et al.}, 2011b), which suggest a funnel flow or disk wind origin rather than an equatorial boundary layer. 

A strong outburst ($\Delta$V$\sim$5\,mag) was intensively monitored in 2008. The source reached peak brightness in 3--4 weeks, remained in a high state for six months with a slight fading, and returned to its initial state within a few more weeks.  
The rapid recovery of EX Lup after the outburst and the similarity between the pre-outburst and post-outburst states suggest that the geometry of the accretion channels did not change between quiescence and outburst, and only the accretion rate varied (e.g., {\it Sicilia-Aguilar et al.}, 2012).

During the outburst, the SED indicates a hot single-temperature blackbody component which emitted 80\%--100\% of the total accretion luminosity ({\it Juh\'asz et al.}, 2012). Strong correlation is also found between the decreasing optical and X-ray fluxes, while UV and soft X-rays are associated with accretion shocks (see Sect.~4.5). From CO fundamental band emission lines, {\it Goto et al.}, (2011) identified a broad-line component that was highly excited, and decreased as the source faded. This gas component likely orbited the star at 0.04--0.4\,AU, implying that it is the inner 0.4\,AU of the disk that became involved in the outburst, a region coinciding with the optically thin, but gas-rich, inner hole. Furthermore,  {\it Juh\'asz et al.}, (2012) concluded, mainly on the basis of accretion timescales, that thermal instability was likely not the triggering mechanism of the 2008 outburst. It remains an open question whether  the inner disk hole, whose radius is larger than the dust sublimation radius, is related to the eruptive phenomenon.

{\it Spitzer} spectra obtained at peak light displayed features of crystalline forsterite, while such crystals were not detected in quiescence (Fig.~\ref{fig:exlupsil}; {\it \'Abrah\'am et al.}, 2009). The crystals were likely produced through thermal annealing in the surface layer of the inner disk by heat from the outburst, a scenario that could produce raw material for primitive comets.  {\it Juh\'asz et al.}, (2012) presented additional multi-epoch {\it Spitzer} spectra, and showed that the strength of the crystalline bands  increased after the end of the outburst, but six months later the crystallinity decreased. Although vertical mixing in the disk would be a potential explanation, fast radial transport of crystals (e.g., by stellar/disk winds) was preferred.  The outburst also influenced the gas chemistry of the disk: {\it Banzatti et al.}, (2012ab) found that the H$_2$O and OH line fluxes increased, new OH, H$_2$, and H\,I transitions were detected, and organics were no longer seen.

In summary, the EX Lup outburst indicated that its enhanced accretion  probably proceeded through magnetospheric accretion channels which were present also in quiescence but delivered less material onto the star. 

\begin{figure*}[!t]
 \epsscale{.95}
 \plotone{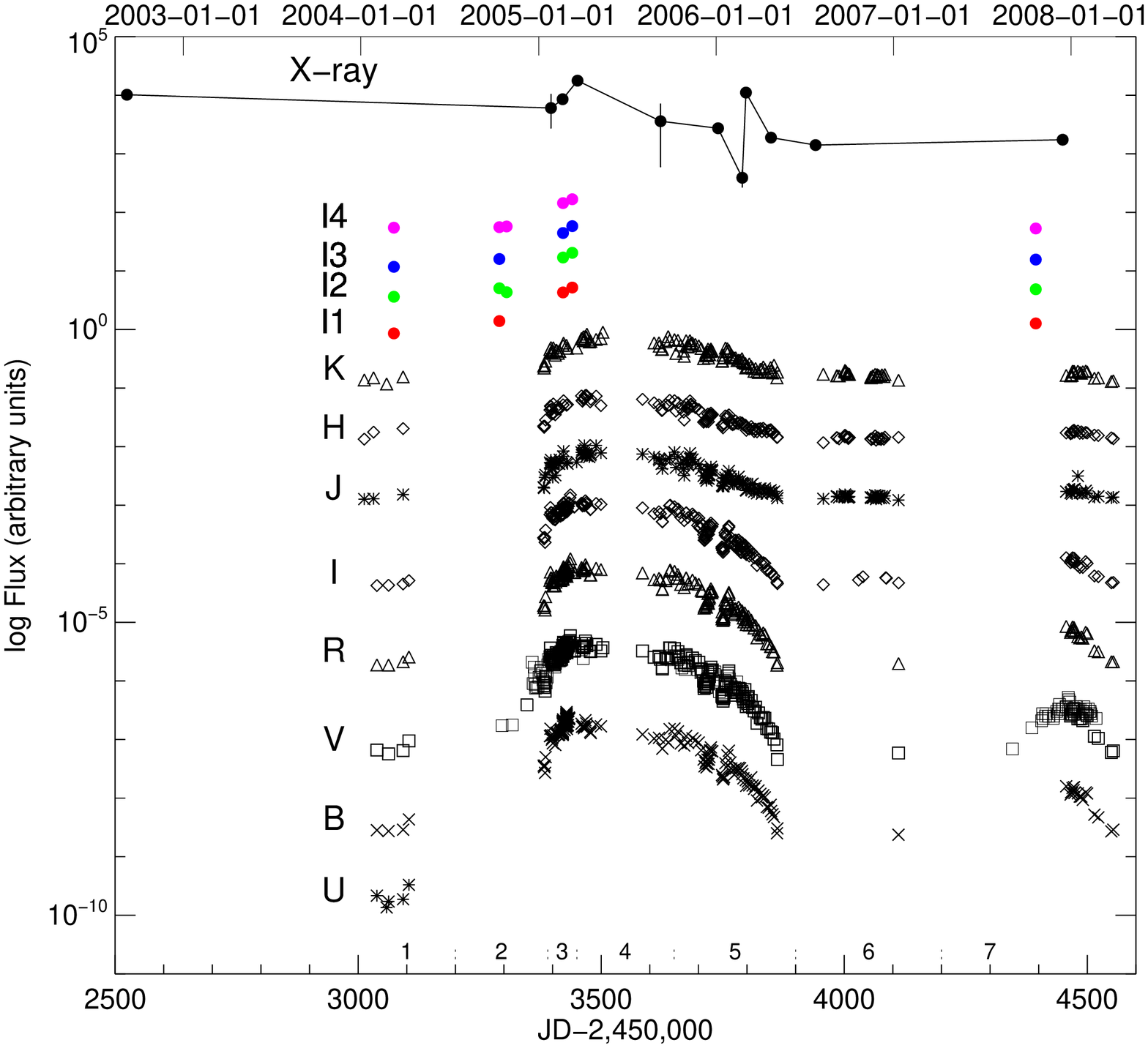}
 \hskip 2mm
 \epsscale{.95}
  \plotone{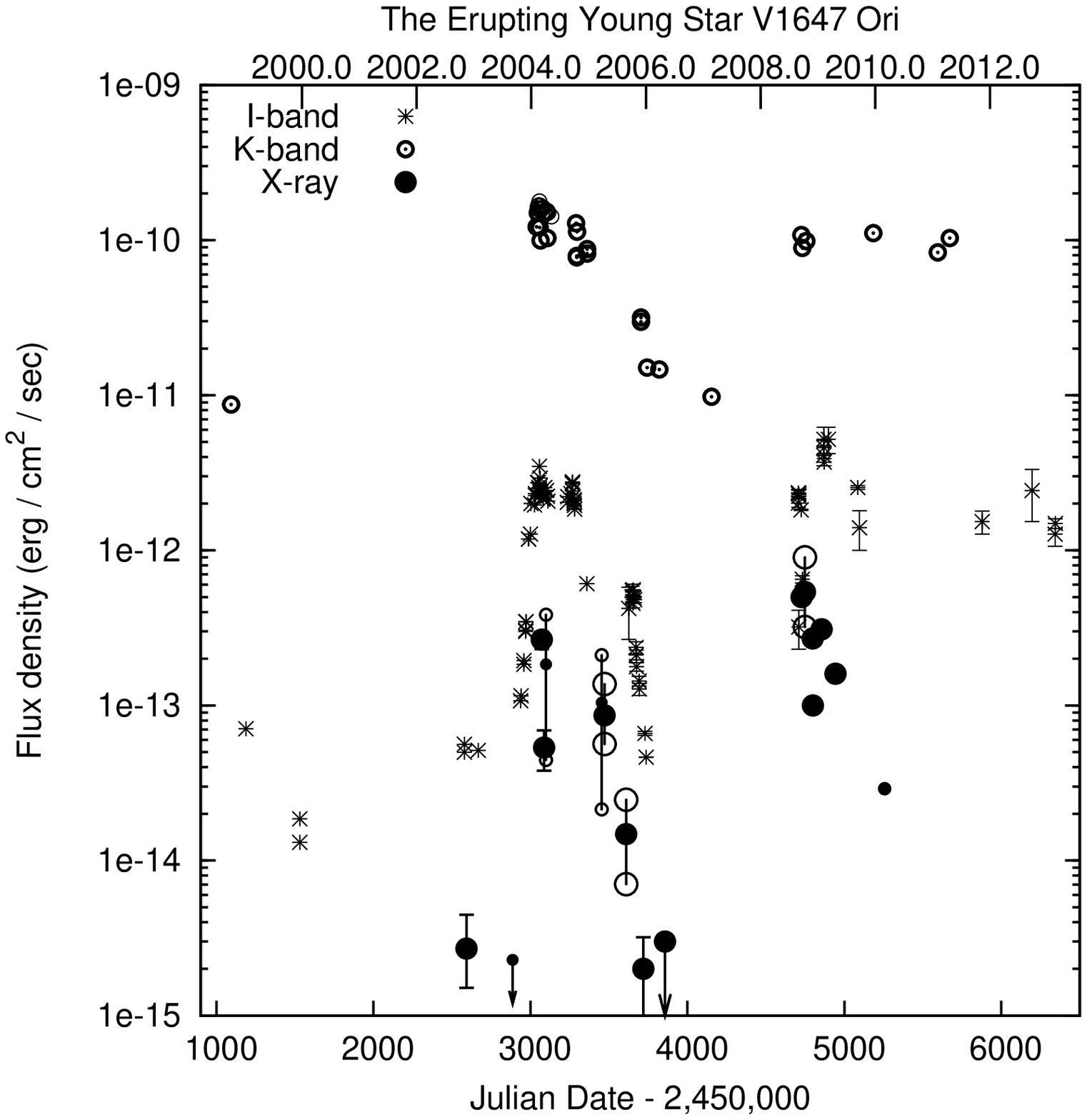}
  \caption{\small Optical, infrared, and X-ray light curves of V1118 Ori (left; from {\it Audard et al.}, 2010; reproduced with permission from Astronomy \& Astrophysics, \copyright ESO) and V1647 Ori (right; M. Richmond, priv.~comm.; adapted from {\it Teets et al.}, 2011). The light curves show that X-rays follow the accretion event, albeit with a weak flux increase in V1118 Ori but a strong increase in V1647 Ori.\label{fig:xrayoptirlc}}
  \end{figure*}

\textit{2.9.2. The classical EXor V1118 Ori.}
The classical EXor V1118 Ori has displayed several outbursts ({\it Herbig et al.}, 2008 and references therein). {\it Parsamian et al.}, (2002) showed that optical spectra taken during the 1989 and 1992-1994 outburst were similar to those of TTS with strong H and Ca~\textsc{ii} lines. The star is a close binary ($0\farcs 18$) with similar fluxes in H$\alpha$ ({\it Reipurth et al.}, 2007b). V1118 Ori was followed during its 2005-2006 outburst in the optical, infrared, and X-rays (Fig.~\ref{fig:xrayoptirlc}; {\it Audard et al.}, 2005, 2010; {\it Lorenzetti et al.}, 2006, 2007) until it returned to quiescence in 2008. The X-ray results are described in Sect.~2.10. The optical and near-infrared emission at the peak of the outburst was dominated by a hotspot ({\it Audard et al.}, 2010; see also {\it Lorenzetti et al.}, 2012), whereas disk emission dominated in the mid-infrared. Star+disk+hotspot models suggested that the mass accretion rate increased from quiescence to peak of outburst from $2.5 \times 10^{-7}$ to $1.0 \times 10^{-6}~M_\sun$~yr$^{-1}$ ({\it Audard et al.}, 2010), together with a significant increase in fractional area of the hotspot. {\it Lorenzetti et al.}, (2012) used a different approach and fitted the difference of the outburst and quiescent SEDs with a black body, obtaining a temperature of $\sim 4500$~K with a radius for the emitting region (assumed to be a uniform disk) of 0.01~AU. $I$-band polarimetry indicated that V1118 Ori is polarized at the level of 2.5\%, with higher and more variable values observed in quiescence, suggesting that the spotted and magnetized photosphere can be seen once the envelope partially disappears ({\it Lorenzetti et al.}, 2007). Color-color diagrams showed variations but no signature of reddening caused by circumstellar matter ({\it Audard et al.}, 2010), consistent with the low extinction ($A_V \leq 2$) found during all activity phases by {\it Lorenzetti et al.}, (2006). {\it Spitzer} spectra showed a slight increase in flux of the crystalline feature at the peak of the outburst compared to post-outburst, but no variation in shape that would indicate a change in grain population. From the CO bandhead emission observed during the declining phase, {\it Lorenzetti et al.}, (2006)  derived a mass loss rate of $(3-8) \times 10^{-8}~M_\sun$~yr$^{-1}$. A  spectrum taken after the outburst detected no emission lines and 2.3~$\mu$m CO in {\it absorption}. {\it Herbig}, (2008) also reported a switch from emission during outburst to absorption in quiescence for Li~\textsc{i}  and K~\textsc{i} lines, and wind emission in H$\alpha$ in the decay phase but a symmetric profile after the outburst. Strong wind losses have, thus, likely  occurred only transiently.

\textit{2.9.3. A new outbursting source: V1647 Ori.}
The outburst of V1647 Ori was discovered in January 2004, illuminating a surrounding nebula. Archival data showed that the source was previously in outburst for 5-20 months in 1966--1967 ({\it Aspin et al.}, 2006). The start of the $21^{\rm st}$ century outburst  occurred in late 2003, leading to an increase of $\sim$5~mag in the $I$-band in 4 months ({\it Brice{\~n}o et al.}, 2004).  Pre-outburst data indicated a flat-spectrum source with an estimated pre-outburst bolometric luminosity of $\approx 3.5-5.6$ \UNITSOLARLUMI\  and age of 0.4~Myr, typical of CTTS ({\it {\'A}brah{\'a}m et al.}, 2004a, see also {\it Andrews et al.,} 2004). The luminosity increased during the outburst by a factor of 10--15 ({\it Andrews et al.}, 2004; {\it Muzerolle et al.}, 2005; {\it Aspin}, 2011b). The peculiarity of V1647 Ori is that its  SED differs from the classical EXors, and more closely resembles those of FUors, with strong extinction (\AV $\sim$6~mag, {\it Ojha et al.}, 2005). The circumstellar mass, however, is typical of TTS ({\it Andrews et al.}, 2004; {\it {\'A}brah{\'a}m et al.}, 2004a; {\it   Tsukagoshi et al.}, 2005).  

During the outburst, the near-infrared color indices of V1647 Ori shifted to bluer colors along the reddening vector, mainly due to intrinsic brightening and partly to decreasing extinction  ({\it Reipurth and Aspin}, 2004a; {\it McGehee et al.}, 2004; {\it Acosta-Pulido et al.}, 2007).  The mass accretion rate during the outburst was estimated at a few $10^{-6}$ to $10^{-5}$~\msunyr\  ({\it Muzerolle et al.}, 2005; {\it Aspin}, 2011b; {\it Mosoni et al.}, 2013). CO bandhead emission was detected along with strong H and He emission lines with P~Cyg profiles, indicating a strong wind with velocities up to 600~\kms, together with ice absorption from H$_2$O and CO and in Na\,I D ({\it Reipurth and Aspin}, 2004a; {\it Vacca et al.}, 2004; {\it Walter et al.}, 2004). {\it Rettig et al.}, (2005) showed  that the CO  emission was hot (2,500~K) and attributed the emission to the accretion zone of the inner disk. They also detected narrow CO absorption lines superimposed on the low-$J$ emission lines  that originated in a foreground cold (18~K) cloud. The ices are unprocessed with a temperature of $\sim$20~K, also indicating  a cloud origin.   In later spectra taken in 2004--2005, evidence for a slow decline in brightness was measured, including a decline of the hot, fast wind (measured from the He~\textsc{i} 1.083$\,\mu$m absorption strength), and of the temperature of the inner disk  ({\it Gibb et al.}, 2006; {\it Ojha et al.}, 2006). From radio data taken in early 2005, {\it Vig et al.}, (2006)  proposed a homogeneous H~\textsc{ii} region to explain the radio, H$\alpha$, and X-ray measurements. Further X-ray results are described in Sect.~2.10.

In late 2005, V1647 Ori suddenly returned to quiescence over a period of a few months ({\it K{\'o}sp{\'a}l et al.}, 2005; {\it Chochol et al.}, 2006; {\it Acosta-Pulido et al.}, 2007; {\it Aspin et al.}, 2008). The He~\textsc{i} 1.083$\,\mu$m line  was then observed in {\it emission}, in contrast to the outburst phase ({\it Acosta-Pulido et al.}, 2007).  Blueshifted CO absorption lines (30 \kms, 700~K) appeared in 2006 Feb superimposed on the previously observed CO emission lines, but were not reobserved in the spectra obtained in 2006 Dec and 2007 Feb ({\it Brittain et al.}, 2007; {\it Aspin et al.}, 2009b). This transient post-outburst outflow was possibly launched by the outward motion of the magnetospheric radius during the rapid decrease of the accretion rate. In 2006 Jan, {\it Fedele et al.}, (2007) also detected   forbidden emission lines indicative of shocked gas likely produced by a HH-like object driven by V1647 Ori. 

Mid-infrared interferometric data obtained during outburst revealed that the emitting region was extended (7~AU at 10~$\mu$m), no close companion could be seen, and the 8--13~$\mu$m spectrum exhibited no strong spectral features ({\it {\'A}brah{\'a}m et al.}, 2006).   Using a disk+envelope geometry and varying the mass accretion rate from (1.6--$7)\times 10^{-6}$ \UNITSOLARMASSYEAR, {\it Mosoni et al.}, (2013) reproduced the SEDs taken at different epochs during the outburst. The models suggested an {\it increase} in the inner radius of the disk/envelope at the beginning of the eruption. However, the system was more resolved at the later epoch when the outbursting region was already shrinking, indicating either that the envelope inner radius suddenly increased at late stages of the outburst, or that the fading of the central source was not immediately followed by the fading of the outer regions.

V1647 Ori returned to the spotlight when a second outburst was reported in 2008 Aug--Sep ({\it Aspin et al.}, 2009c). The source photometry, accretion rate, luminosity, and morphology were similar to those seen after the onset of the previous outburst.  However, CO overtone emission was not detected, despite being seen shortly after the onset of the 2003 outburst. {\it Aspin et al.}, (2009c) concluded that the quiescent period between these two outbursts was due to a partial decline in the accretion rate (factor of 2--3) and reformation of dust in the immediate circumstellar environment. They argued that the 2008 outburst was caused by the combination of enhanced accretion and sublimation of dust due to this energetic event. In the high state of the second outburst, the H$\alpha$ and  Ca~\textsc{ii}  lines did not change remarkably ({\it Aspin}, 2011b). The CO overtone bandhead was still not detected, while the water vapor absorption remained strong. {\it Brittain et al.}, (2010) further found that  the temperature of the CO emission had varied with the accretion rate, and showed  a direct relation between the Br$\gamma$ luminosity and line full width at half maximum and the source brightness, indicating that the accretion rate had driven the variability.

In view of the duration of the outburst, its recurrence, and the various spectroscopic features observed before, during, and after the outburst, V1647 Ori does not fit well either with classical EXors or  FUors;  it may instead be  an intermediate case. In fact, {\it Aspin et al.}, (2009b) presented a spectrum of V1647 Ori, taken in quiescence, that does not resemble those of late-type standards, Class I protostars, or EXors observed in quiescence, but  does show considerable correspondence with several classical FUors observed {\it during} their outbursts.  Given that V1647 Ori shares some characteristics of both FUors and EXors, they proposed the existence of a ``continuum'' of eruption  characteristics rather than two distinct classes.  

\bigskip
\noindent
\textbf{2.10. High-Energy Processes}
\bigskip

Strong X-ray emission is characteristic of young stars ({\it Feigelson and Montmerle}, 1999), but  our knowledge of the high-energy behavior of eruptive young stars is limited to just a few examples observed over the past decade or so. X-rays probe high-energy processes, such as coronal activity and accretion shocks, and they are an important source of ionization and heating of accretion disk atmospheres, and thus they may influence disk chemistry and strengthen the coupling of disk gas to the stellar magnetic field. 

\textit{2.10.1. X-rays from Classical FUors.}
The first X-ray spectrum of FU Ori obtained with {\it XMM-Newton} appeared quite unusual. The emission consisted of a hot plasma  viewed through very high absorption --- much higher than anticipated based on $A_V$ --- and a cooler component  seen through ten times lower absorption  ({\it Skinner et al.}, 2006).  Subarcsecond X-ray imaging by {\it Chandra} subsequently provided an explanation for the sharply contrasting absorbing columns ({\it Skinner et al.}, 2010): the high-temperature component was positionally coincident with FU Ori, while the centroid of the cooler component was offset from FU Ori and displaced toward the infrared companion FU Ori S. The ``excess'' absorption toward FU Ori is likely a result of accreting gas, FU Ori's powerful wind, or both.  Time-variability in the hot component --- implying a magnetic origin --- was detected, but no such variability was seen in the fainter, less-absorbed, cooler component. The classical FUor V1735 Cyg was bright ($\log L_{X} \approx 31.0$~ergs s$^{-1}$) and displayed very hot ($T > 70$ MK), heavily-absorbed plasma but no cool plasma, possibly because the latter is more susceptible to absorption by intervening cool gas.  In contrast, V1057 Cyg and V1515 Cyg both escaped detection ({\it Skinner et al.}, 2009). Interestingly a faint soft X-ray jet was detected with {\it Chandra} in Z CMa after the outburst  ({\it Stelzer et al.}, 2009), with a position angle consistent with that of the micro-jet launched by the FUor component of this close binary ({\it Whelan et al.}, 2010). The jet was not detected in a pre-outburst observation, suggesting that mass ejection occurred. Clearly more observations are needed to characterize the X-ray properties of the FUor class as a whole. 

\textit{2.10.2. X-rays from Classical EXors.}
The eruptions of  V1118 Ori in 2005-2006 and EX Lup in 2008 were both monitored in X-rays. {\it Teets et al.}, (2012) followed the EX Lup outburst beginning two months after the outburst onset until just after its conclusion, while {\it Audard et al.}, (2005, 2010) and {\it  Lorenzetti et al.}, (2006) followed V1118 Ori until it returned in quiescence in 2008 (Fig.~\ref{fig:xrayoptirlc}). In both outbursts, there were strong correlations between the decreasing optical and X-ray fluxes following the peak of the optical outburst, suggesting a relation with the accretion rate. However --- in contrast to V1647 Ori (see below) --- the X-ray flux increased only mildly in both cases, with a decrease in flux after outburst, and relatively cool plasma  temperatures ($5-8$~MK)  were observed. The temperature in V1118 Ori contrasted with a serendipitous pre-outburst observation in 2002 that showed a dominant 25~MK plasma ({\it Audard et al.}, 2005). The plasma temperature then gradually returned to higher values in later phases of the outburst ({\it Audard et al.}, 2010), in similar fashion to EX Lup ({\it Teets et al.}, 2012) --- although the cool plasma was still detected before the end of the 2008 EX Lup outburst ({\it Grosso et al.}, 2010). 

One possible explanation for the anticorrelation between optical/infrared flux and X-ray temperature observed for both V1118 Ori and EX Lup is that their soft X-ray outputs were enhanced during eruption, due to emission arising in accretion shocks (along with possible changes in magnetospheric configurations). In the case of EX Lup, this hypothesis is consistent with the strong variability in ultraviolet emission detected by {\it XMM-Newton}, which was likely due to accretion spots; in addition, the X-ray emission at energies above 1.5 keV showed far stronger photoelectric absorption than the cool plasma, suggesting the coronal emission was subject to absorption by overlying, high-density gas in accretion streams  ({\it Grosso et al.}, 2010). 

\textit{2.10.3. X-rays from V1647 Ori.}
{\it Kastner et al.}, (2004, 2006) followed the entire 2003--2005 duration of V1647 Ori's outburst with {\it Chandra}, including a serendipitous pre-outburst observation. The X-ray flux from V1647 Ori closely tracked the rise and fall of its optical-infrared flux (Fig.~\ref{fig:xrayoptirlc}), thereby providing definitive evidence for the production of high-energy emission via star-disk interactions. The monitoring data also indicated hardening of the X-ray emission during outburst. Further monitoring of V1647 Ori in 2008--2009 --- i.e., soon after the onset of its second outburst --- established the striking similarity between the two outbursts in terms of X-ray/near-infrared flux ratio and X-ray spectral hardness ({\it Teets et al.}, 2011).

Deep {\it XMM-Newton} and {\it Suzaku} exposures were obtained during V1647 Ori's first and second outbursts, respectively ({\it Grosso et al.}, 2005; {\it Hamaguchi et al.}, 2010). The observations showed warm (10~MK) and hot (50~MK) plasmas, the former indicative of high-density plasma associated with small magnetic loops around coronally active stars ({\it  Preibisch et al.}, 2005), the latter consistent with magnetic reconnection events. Strong fluorescent Fe K$\alpha$ emission was detected, providing evidence for irradiation of neutral material residing either in the circumstellar disk or at the stellar surface. The comparison of $A_V$ ($\approx 10$~mag) with the hydrogen column density  ($N_{\rm H} \sim 4\times 10^{22}$~cm$^{-2}$) points out a significant excess of X-ray absorption from relatively dust-free material.

Via time-series analysis, {\it Hamaguchi et al.}, (2012) established that the puzzling short-term (hours-timescale) X-ray variability of V1647 Ori was due to rotational modulation. The $\sim$1 day X-ray period corresponds to rotation at near-breakup rotation speed.  {\it Hamaguchi et al.}, (2012)  further demonstrated that a model consisting of two pancake-shaped hot spots of high plasma density ($\ge5\times10^{10}$ cm$^{-3}$), located at or near the stellar surface, reproduced well the X-ray rotational modulation signature. Under the assumption that the hot spots are generated via magnetic reconnection activity, the stability of the X-ray light curve during the course of two accretion-related outbursts further suggests that the star-disk magnetic configuration of V1647 Ori has remained relatively unchanged over timescales of years.

\bigskip
\centerline{\textbf{3. THEORY}}
\bigskip
\noindent
\textbf{3.1. SED Fits and Accretion Disk Origin}
\bigskip

\begin{figure}[!t]
 \epsscale{1.0}
 \plotone{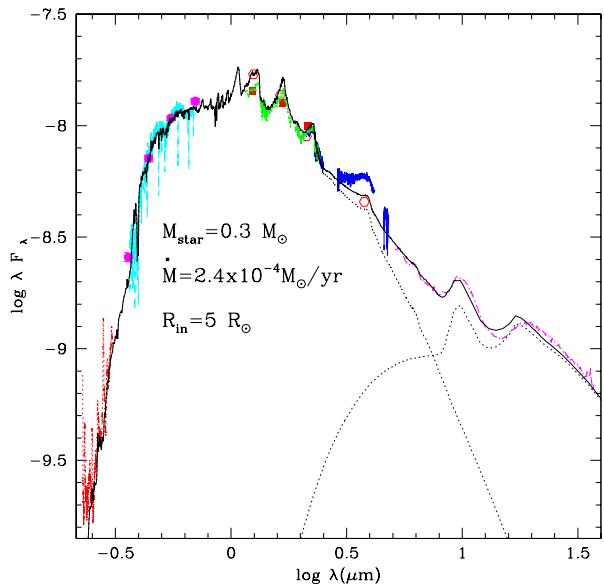}
 \caption{\small Observed SED of FU Ori (solid line) fitted with 
a steady accretion disk model with contributions from the inner hot disk (light dotted curve) and the flared outer disk (light dashed curve). Adapted from {\it Zhu et al.}, (2008).\label{fig:zhufuori}}  
 \end{figure}

SED fitting based on theoretical models is a powerful too in understanding the origin of these outbursting objects.
The most successful model to explain the peculiarities of FUors was proposed by {\it Hartmann
and Kenyon}, (1985, 1987ab) and {\it Kenyon et al.}, (1988), who suggested that the SED was dominated by a protostellar disk accreting at a high rate ($\dot{M}\approx 10^{-5}-10^{-4}$~\msunyr). Within this picture, it is easy to account for the main peculiarities observed in FUor
spectra, such as the changing supergiant spectral types from optical to infrared, and the double-peaked line profiles,
with peak separation decreasing with increasing wavelength  ({\it Hartmann and Kenyon}, 1996).

Self-consistent disk atmospheric models including  the accretion process and the radiative transfer in the disk atmosphere are needed to  constrain the SED and, in particular, disk properties (e.g., {\it Calvet et al.}, 1991ab; {\it Popham et al.}, 1996). {\it Zhu et al.}, (2007, 2008, 2009c) extended the first models with a more complete opacity database ({\it Kurucz et al.}, 1974). This simple steady accretion disk model could fit FU Ori's SED (Fig.~\ref{fig:zhufuori}; {\it Zhu et al.}, 2007, 2008) and confirmed the Keplerian rotation of FU Ori's disk ({\it Zhu et al.}, 2009a). The derived size of the high accretion rate inner disk was from  5 $R_{\odot}$ to $\sim$0.5--1 AU, beyond which is a passively heated outer disk.  The outer value is consistent with the hot disk size found by {\it Eisner and Hillenbrand}, (2011). Incidentally, such a size leads to a viscosity parameter $\alpha \sim 0.02 - 0.2$ in  outburst ({\it Zhu et al.}, 2007).

Different from this traditional steady viscous disk approach, {\it Lodato and Bertin}, (2001) argued that FUor disks might be gravitationally unstable. The radial structure of such a gravitationally unstable disk should be significantly different from a non-self-gravitating one. They contended that the observed flat SED in the infrared might be related to a combination of extra heating induced by non-local dissipation of large-scale spiral structures and by the possible departure from a purely Keplerian rotation, if the disk mass is a sizeable fraction of the central object (see also {\it Adams et al.}, 1988; {\it Bertin and Lodato}, 1999). {\it Lodato and Bertin}, (2003) then investigated how such effects would modify the shape of the double-peaked line profiles.

\bigskip
\noindent
\textbf{3.2. Origin of the Outburst}
\bigskip

A closely related question for the disk accretion model pertains to the origin of the outburst. Two main schools of thought have been proposed to explain the triggering of FUor outbursts: 1) disk instability and 2) perturbation of the disk by an external body. 

Thermal instability is due to the thermal runaway process when the disk temperature reaches
the hydrogen ionization temperature. In detail, a hydrostatic viscous disk can 
be thermally stable only if the opacity changes slowly with the  temperature. When the
disk temperature reaches the hydrogen ionization temperature, the opacity increases
dramatically with the temperature ($\kappa\sim T^{10}$). A slight temperature increase will
cause a significant amount of energy trapping in the disk and the disk temperature continues to
rise, which leads to the thermal runaway. The latter ends when the disk is fully ionized
at $\sim 10^{5}$~K and the opacity changes slowly with temperature again. 
After the thermal runaway, a high disk temperature can lead to a high disk accretion rate since the viscosity $\nu$
is proportional to the disk temperature ($\nu=\alpha c_{s}^{2}/\Omega \propto \alpha T/\Omega$, with $c_s$ the sound speed, viscosity parameter $\alpha$, and $\Omega$ the angular velocity).

The thermal instability model can naturally explain the fast rise time of FU Ori. 
However, since thermal instability needs to be triggered at T$\sim$5000 K, 
the outbursting disk is small with a size of $\sim$20$R_{\odot}$. In order to maintain
the outburst for hundreds of years, the $\alpha$ parameter needs to be as low as 10$^{-3}$
during the outburst. Furthermore, in order to produce enough accretion rate contrast before and during the outburst,
$\alpha$  needs to be 10$^{-4}$ in the quiescent state ({\it Bell and Lin}, 1994; {\it Bell et al.}, 1995).
2-D radiation hydrodynamic simulations have been carried out to study the triggering of thermal instability
in disks and the thermal structure of the boundary layer ({\it Kley and Lin}, 1996, 1999; {\it Okuda et al.}, 1997).
The SED fitting for FU Ori does not require an additional hot boundary layer (Fig.~\ref{fig:zhufuori}), which may suggest that
the boundary layer can be different between FUors and TTS ({\it Popham}, 1996).

\begin{figure}[!t]
 \epsscale{1.0}
 \plotone{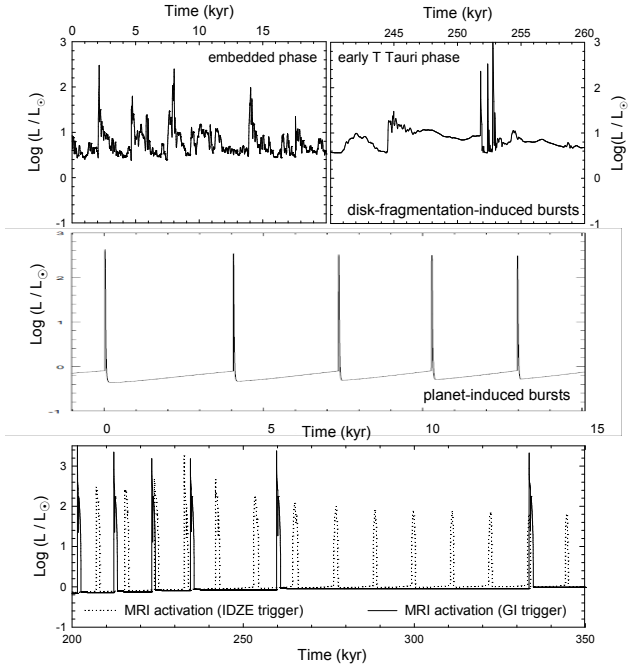}
 \caption{\small {\it Top: }  Total (accretion plus photospheric) luminosity in
the disk fragmentation model showing typical outbursts in the embedded phase
of star formation (left) and in the early T Tauri phase (right). 
{\it Middle:} Bolometric lightcurves for the planet induced thermal instability model. 
Within this model the recurrence time between outbursts is reduced as time increases.
{\it Bottom:} Total luminosity for the MRI thermally activation models. 
The solid curves are from models where MRI is triggered by GI, while
the dotted curves are from models where MRI is activated at the inner dead zone
edge (IDZE) due to a non-zero dead zone viscosity.\label{fig:bursts1}
}  
 \end{figure}

\begin{figure}[!t]
 \epsscale{1.0}
 \plotone{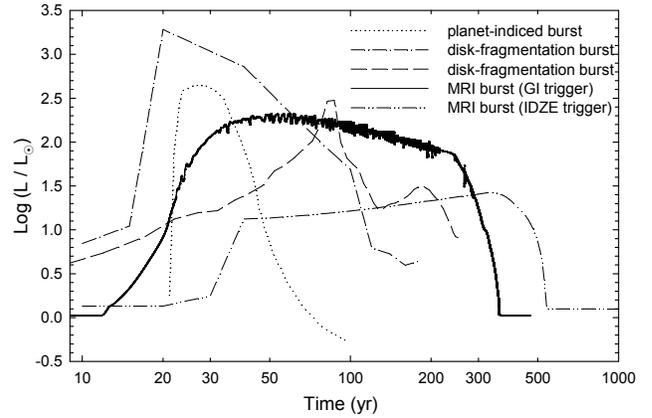}
 \caption{\small Time evolution of individual luminosity outbursts in different
burst-triggering models.  The zero-time is chosen arbitrarily to highlight distinct models.
The two distinct outburst curves in the disk fragmentation model stem from
the state of the fragment when accreted onto the star. Tidally smeared
fragments give rise to a slowly rising curve (predominantly, in the embedded phase), while a sharply rising curve is produced by weakly
perturbed fragments in the early T Tauri phase.\label{fig:bursts2}
}  
 \end{figure}

Several models have been put forward to cope with the limitations of the
standard thermal instability model. These include variations on the
thermal instability model itself, or an altogether new trigger mechanism,
for example associated with the onset of the magneto-rotational or
gravitational instability. They are discussed in detail  below. 
A comparison of the light curves produced by these different models can 
be found in Figs.~\ref{fig:bursts1} and \ref{fig:bursts2}, which show their long-term evolution (over a 
period of thousands of years) and the detailed evolution of a single 
outburst, respectively.

{\it 3.2.1. Thermal instabilities induced by a planet.}
In thermal instability models, it has been noted that, in the absence of a trigger, the instability would first occur in the innermost parts of the disk and then propagate inside-out, in a ``snowplough" fashion ({\it Lin et al.}, 1985), decelerating as it propagates out and thus leading to a slow rise in the light curve (as observed in the case of V1515 Cyg). A fast rise (as observed in FU Ori and V1057 Cyg) is produced only if the outburst is ``triggered" somehow far from the disk inner edge, so that the instability can propagate outside-in, in an ``avalanche" fashion. {\it Clarke et al.}, (1990) have confirmed this behavior by including an {\it ad hoc} density perturbation to produce triggered outbursts. The latter approach has been developed initially by {\it Clarke and Syer}, (1996) and then by {\it Lodato and Clarke}, (2004), who considered outbursts triggered by a massive planet. During quiescence the planet opens up a gap in the disk. For sufficiently massive planets, the resulting Type II migration is slow and the inner disk is rapidly emptied out (in what would resemble a transitional disk), leading to a steepening of the density profile in the outer disk, that can trigger the thermal-viscous instability at the outer gap edge. Observationally, there are a number of clues that indicate the presence of a planet in FUor disks.  {\it Clarke and Armitage}, (2003) have suggested that a planet embedded in the disk of a FUor would lead to a clear spectroscopic signature in the form of a periodic modulation of the double-peaked line profiles on the orbital period of the planet. Such periodic modulations of the line profiles have been observed only for the fast rise FUors (FU Ori and V1057 Cyg), with a period of $\sim$ 3 days ({\it Herbig et al.}, 2003), corresponding to a semi-major axis of 7-10 $R_{\odot}$ (see also {\it Powell et al.}, 2012).

The results of {\it Lodato and Clarke}, (2004) confirm that a planet with a mass of a few Jupiter masses can produce a fast rise outburst (with a rise time of the order of one year, as observed for FU Ori and V1057 Cyg). The issue with this class of models is the duration of the outburst, predicted to be of the order of $\sim 50$ years by such models, with a relatively shallow dependence on the planet mass. This timescale is set by the viscous time on the ionized branch of the stability curve at the outermost location reached by the instability front, that is generally confined within $\sim 50R_{\odot}$, thus making this timescale too small, unless $\alpha$ is assumed to be unrealistically low in the upper branch ($\alpha\sim 10^{-4}$). 

{\it Clarke et al.}, (2005) compared planet-triggered outburst models to a long-term monitoring campaign of FU Ori and V1057 Cyg optical light curves. The time-dependent models were able to match the color evolution of the light curves much better than a simple series of steady-state models with varying $\dot{M}$. {\it Clarke et al.}, (2005) also discussed the sudden luminosity dips observed for V1057 Cyg at the end of the outburst and for V1515 Cyg, while FU Ori does not appear to show any such behavior. This non-periodic variability was interpreted as a consequence of the interaction of a disk wind with the infalling envelope. Numerical simulations show that powerful winds are able to push away the infalling envelope to large radii clearing up our view to the inner disk, while for less powerful ones the envelope ``crushes down'' the wind occulting the inner disk. If the wind strength is proportional to the accretion rate in the inner disk, such models can explain the observed behavior.

The interaction between the disk and a planet can be much more complex if there is mass transfer between the planet and the star.  {\it Nayakshin and Lodato}, (2013) considered the case of a few Jupiter masses inflated planet, migrating within a  disk. When the planet opens up a gap, mass transfer between the planet and the star leads to the planet losing mass but retaining angular momentum, thus migrating out and switching off the mass transfer. Conversely, if the gap is only partially open, a runaway configuration can occur where the planet migrates further in, enhancing the mass loss rate and giving rise to a powerful flare. This mechanism gives rise to variability on several timescales and of different amplitudes depending on parameters.

{\it 3.2.2. A combination of gravitational instability and magnetorotational instability.}
Both magnetorotational instability (MRI, {\it Balbus and Hawley}, 1998 and references therein; see also the accompanying chapter by {\it Turner et al.}) and gravitational instability (GI, see {\it Durisen et al.}, 2007 and references therein) are promising mechanisms to transfer angular momentum in disks. However, MRI only operates when the disk reaches enough ionization (that is, in the hotter parts of the disk), and GI only operates when the disk is sufficiently cold to become gravitationally unstable, according to the classical criterion ({\it Toomre}, 1964):
\begin{equation}
Q=\frac{c_{\rm s}\kappa}{\pi G\Sigma}\approx \frac{c_{\rm s}\Omega}{\pi G\Sigma} \lesssim 1,
\end{equation}
where $Q$ is the Toomre stability criterion, $\Sigma$ is the surface density and the epicyclic frequency $\kappa$ can be approximated with the angular frequency $\Omega$ for a quasi-Keplerian disk. A realistic protoplanetary disk is likely to have a complicated
accretion structure: the inner disk is MRI active due to the thermal ionization, the region
between $\sim$AU to tens of AU has a layered accretion structure with a MRI
stable dead zone, and the outer region can be MRI active due to cosmic/X-ray ionization. The outer disk
can also be gravitationally unstable when the disk is at the infall phase with a significant amount of mass loading.

This complicated structure is unlikely to maintain a steady accretion  ({\it Gammie}, 1996; {\it Armitage et al.}, 2001;
{\it Zhu et al.}, 2009b, 2010a; {\it Martin et al.}, 2012ab). The outer disk transfers mass to the inner disk either due to 
GI or due to the layered accretion. With more and more mass shoveled to the inner disk, the inner disk becomes gravitationally unstable and continues to  transfer mass to small radii. GI can also heat up the disk. Since the electron fraction in protoplanetary disks  depends nearly exponentially on temperature due to the collisional ionization of potassium ({\it Gammie}, 1996; {\it Umebayashi}, 1983), the gaseous disk will be well coupled to the magnetic field when $T \sim 1000$~K and MRI starts to operate. This sudden activation of MRI leads to disk outbursts. 
During the outburst, a thermal instability should also be triggered at the inner disk as a by-product. Two dimensional
radiation hydrodynamic simulations have been carried out to verify such mechanisms ({\it Zhu et al.}, 2009b).

This MRI + GI mechanism shares similarities with the traditional thermal instability mechanism. For a given disk surface density, the disk has two possible structures: 1) MRI inactive, and 2) MRI active. The switch from one to another occurs
at $T \sim 1000$~K and leads to outbursts ({\it Martin and Lubow}, 2011, 2013). In the MRI+GI picture,
GI triggers the switch. However, if there are other mechanisms to heat up the disk and trigger the switch,
the disk can also experience outbursts. For example, {\it Bae et al.}, (2013) found that  the energy diffusion radially 
from the inner MRI active disk to the dead zone can trigger the switch and lead to outbursts, although these outbursts are weaker and shorter than the MRI+GI case. This may have implications for weaker FUors or EXors.

After the envelope infall stage, the layered accretion can still pile up mass in the dead zone and leads to
disk outbursts ({\it Zhu et al.}, 2010b), which is consistent with the fact that some FUors are in the T Tau phase.
However, this massive dead zone would persist throughout the whole T Tauri phase and should be easily
observed by ALMA.

{\it 3.2.3. Disk fragmentation.}
Another accretion burst mechanism is  related to the phenomenon of disk gravitational 
fragmentation and subsequent inward migration of the fragments onto the protostar.
Observations of jets and outflows suggest that the formation 
of protostellar disks often occurs in the very early stage of star formation, 
when a protostar is deeply embedded in a parental cloud core. 

The forming disk becomes gravitationally unstable, if the $Q$-parameter drops below unity. 
A higher initial mass and angular momentum of the parental core both 
favor the formation of  disks with stronger gravitational instability
({\it Kratter et al.}, 2008; {\it Vorobyov}, 2011). 
For relatively long cooling timescales (above a few dynamical timescales) the outcome of the 
instability is to produce a persistent spiral structure ({\it Gammie}, 2001; {\it Lodato and Rice},
2004, 2005; {\it Mej\'{\i}a et al.}, 2005; {\it Boley et al.}, 2006), able to transport angular momentum 
efficiently through the disk ({\it Cossins et al.}, 2009, 2010)
and trigger the MRI burst (Sect.~3.2.2).

In the opposite regime, where the cooling timescale is comparable 
to or shorter than the dynamical timescale, 
these disks can experience gravitational fragmentation and 
form bound fragments with mass ranging from giant planets to very-low-mass stars 
({\it Gammie}, 2001; {\it Johnson and Gammie}, 2003; {\it Rice et al.}, 2005;
{\it Stamatellos and Whitworth}, 2009ab; 
{\it Vorobyov and Basu}, 2010; {\it Vorobyov}, 2013). 

The likelihood of survival of the fragments is, however, low.
GI in the embedded phase is strong, fueled with a continuing infall
of gas from a parent cloud core, and resultant gravitational
and tidal torques are rampant. As a consequence, the fragments tend to be driven 
into the inner disk  and likely onto the protostar due to the loss of 
angular momentum caused by the gravitational interaction with the trailing 
spiral arms ({\it Vorobyov and Basu}, 2006, 2010; {\it Cha and Nayakshin}, 2011; 
{\it Machida et al.}, 2011). The infall of the fragments can trigger mass accretion 
bursts  and associated luminosity 
outbursts that are similar in magnitude to FUor or EXor events 
({\it Vorobyov and Basu}, 2005, 2006, 2010),
during which a notable fraction of the protostellar mass can be accumulated 
({\it Dunham and Vorobyov}, 2012). The protostellar accretion pattern
in the disk fragmentation models is highly variable ({\it Vorobyov}, 2009) 
and is characterized by short-duration ($\la 100-200$~yr) bursts with $\dot{M} \ga 
\mathrm{a~few} \times 10^{-5}$~\msunyr\  
alternated with longer ($ 10^{3}$--$10^{4}$~yr) quiescent periods with 
$\dot{M} \la 10^{-6}$~\msunyr.
The most intense accretion bursts can reach $10^{-3}$~\msunyr.

This burst mechanism is most effective in the embedded stage of disk evolution and
diminishes once the parental core has accreted most of its mass reservoir onto the protostar + disk
system. Only occasional bursts associated with the fragments that happened to survive through
the embedded stage can occur in the T Tauri stage. Moreover, the initial conditions in the parental
core set the likelihood of disk fragmentation and the burst occurrence; the relevant rotational energy
vs.~initial core mass diagram is provided in {\it Basu and Vorobyov}, (2012) and 
{\it Vorobyov}, (2013). 
Magnetic fields and strong background irradiation both act to suppress disk fragmentation
and associated accretion bursts, but are unlikely to quench this phenomenon completely
({\it Vorobyov and Basu}, 2006, 2010).

\textit{3.2.4. External Environment Triggers.}
The accretion and luminosity burst mechanisms considered in the previous sections 
are induced by ``internal'' triggers ({\it Tassis and Mouschovias}, 2005). These triggers  depend on the physical conditions in the system and if they are suppressed, the protostar is likely to accrete in a quiescent manner. Nevertheless, accretion bursts can still be
induced in quiescent systems by the so-called ``external'' triggers, of which a close encounter 
in a binary system was the first proposed candidate ({\it Bonnell and Bastien}, 1992). 
{\it Reipurth and Aspin}, (2004b)  considered the case where dynamical interactions in small-N systems might lead to close encounters. Starting from the fact that a couple of FUors are found to be in a binary system where both stars are FUor, they argued that whatever triggered the FUor eruption in one component is likely to also have triggered the eruption in the other component, and identified the breakup of a small multiple system as a natural common precursor event. Indeed, dynamical interaction in small-N systems can easily result in the formation of a close binary. {\it Reipurth and Aspin}, (2004b)  then argued that the decay of the binary due to interaction with a circumbinary disk may lead to triggering of instabilities in the individual circumstellar disks. A quantitative modeling of such processes, however, is still lacking.

The idea of external triggering received further development by {\it Pfalzner et al.}, (2008) and 
{\it Pfalzner}, (2008), who considered accretion processes in young and dense stellar clusters, choosing
the Orion nebular cluster as representative.
They combined cluster simulations performed with the N-body code and particle simulations 
that described the effect of a close passage on the disk around a young star to determine 
the induced mass accretion. They concluded that the close encounters
can reproduce the accretion bursts typical for the FUor events, albeit with certain limitations.

Close encounters with nearby stars have also been considered as a way to induce fragmentation in a disk that is gravitationally stable, but not fragmenting in isolation ({\it Mayer et al.}, 2005), although the effectiveness of such process is probably limited ({\it Lodato et al.}, 2007).

However, the short rise times are difficult to achieve unless the matter is 
stored somewhere close to the protostar and the inferred duration of FU Ori events requires a rather
high mass ratio between the intruder and the target. 

\bigskip
\noindent
\textbf{3.3.
Episodic Accretion and its Implications}
\bigskip

Episodic accretion has several key implications for star and planet formation and evolution.

{\it 3.3.1 Resolving the luminosity problem in embedded sources.}
In one of the first statistically significant studies of the luminosities of 
embedded protostars, {\it Kenyon et al.}, (1990, 1994) and {\it Kenyon and 
Hartmann}, (1995) found that the observed luminosities of 23 protostars in 
Taurus were much lower than expected from simple theoretical predictions.  
Current samples of  protostars with accurately determined luminosities measure in the hundreds 
(e.g., {\it Evans et al.}, 2009; {\it Enoch et al.}, 
2009; {\it Kryukova et al.}, 2012; {\it Dunham et al.}, 2013; also see the 
accompanying chapter by {\it Dunham et al.}).  While the 
details of these studies differ, they all confirm the existence of the 
luminosity problem and aggravate it by showing that the luminosity 
distribution of protostars extends to even lower luminosities than found by 
the {\it Kenyon et al.} surveys.

As originally proposed by {\it Kenyon and Hartmann}, (1995), one possible 
resolution to the ``luminosity problem" is episodic accretion.  If a significant 
fraction of the total accretion onto a star occurs in short-lived bursts, 
most protostars will be observed during periods of accretion below the 
mean accretion rate, thus most protostars will emit less accretion luminosity 
than expected assuming constant accretion at the mean rate.  
To test the ability and necessity of episodic accretion for resolving the 
luminosity problem, {\it Dunham et al.}, (2010) modified existing evolutionary 
models of collapsing cores first published by {\it Young and Evans}, (2005) 
and showed that a very simple implementation of episodic accretion into 
the accretion model was capable of resolving the luminosity problem and 
matching the observed protostellar luminosity distribution.   {\it Offner and McKee}, (2011) also found
that episodic accretion can contribute a significant fraction of the stellar mass.
{\it Dunham and Vorobyov}, (2012) revisited this topic with more sophisticated 
evolutionary models incorporating the exact time evolution of the accretion 
process predicted by hydrodynamical simulations, and confirmed these findings.  
However,  these results only demonstrate that episodic 
accretion is a {\it possible} solution to the luminosity problem.

{\it 3.3.2. Impact on disk fragmentation.}
Stellar irradiation is known to have a profound impact on the gravitational 
stability of protostellar disks ({\it Stamatellos and Whitworth}, 2009ab; 
{\it Offner et al.}, 2009, {\it Rice et al.}, 2011). Flared 
disks can intercept a notable fraction of the stellar UV and X-ray flux, 
which first heats dust and then gas via collisions with dust grains. The net 
effect is an increase in the gas/dust temperature leading to disk 
stabilization.  X-rays can also ionize gas.
SPH simulations by {\it Stamatellos et al.}, (2011, 2012) demonstrate that quiescent 
periods between the luminosity bursts can be sufficiently long for the disk 
to cool and fragment. Thus, episodic accretion can enable 
the formation of low-mass stars, brown dwarfs, and planetary-mass objects by 
disk fragmentation, as confirmed by the recent numerical hydrodynamical 
simulations presented by {\it Vorobyov}, (2013).

{\it 3.3.3. Luminosity spread in young clusters.}
The luminosity spread observed in Hertzsprung–-Russell (HR) diagrams 
of young star-forming regions is a well known feature ({\it Hillenbrand}, 2009), 
but its origin is uncertain. It can be attributed to observational uncertainties,
significant age spread, or yet unknown physical processes. As {\it Baraffe et al.}, (2009)
demonstrated, protostellar evolutionary models assuming episodic accretion are able 
to reproduce the observed luminosity spread for objects with final ages of a few Myr.
In contrast, non-accreting models require an age spread of at least 10~Myr. 
 {\it Baraffe et al.}, (2009), thus, concluded that the observed HR luminosity spread does not stem entirely from an age
spread, but rather from the impact of episodic accretion on
the physical properties of the protostar. This idea was recently called into question by 
{\it Hosokawa et al.}, (2011), who argued that accretion variability
had little effect on the evolution of low-mass protostars with effective temperature below 4000~K.
   
{\it Baraffe et al.}, (2012) noted that {\it Hosokawa et al.}, (2011) considered models with 
the initial protostellar ``seed" mass of 10~$M_{\rm Jup}$. Having provided justification for a smaller initial mass 
of protostars on the order of 1.0~$M_{\rm Jup}$, {\it Baraffe et al.}, (2012) showed that 
the luminosity spread in the HR diagram and the inferred properties of FU Ori events 
(stellar radius, accretion rate) both can be explained by the ``hybrid" accretion scenario with 
variable accretion histories derived from disk fragmentation models (Sect.~3.2.2).
In this accretion scenario, a protostar absorbs no accretion
energy below a threshold accretion rate of $10^{-5}$~\msunyr\  and 20\% of the accretion 
energy above this value.

Several important implications of protostellar evolutionary models with variable accretion 
were emphasized by {\it Baraffe et al.}, (2012).
First, each protostar/brown dwarf experiences its own accretion history
and ends up randomly in the HR diagram at the end of the accretion phase. It is likely
that the concept of a birthline does not apply to low-mass ($<1.0~\msun$) objects.
Moreover, age determinations from ``standard" non-accreting isochrones may 
overestimate the age of young protostars by a factor of several or more.
Finally, inferring masses from the HR diagram using 
isochrones of non-accreting protostars can yield severely incorrect determinations,
possibly overestimating the mass by as much as 40\% or more.

\begin{figure}[!t]
 \epsscale{1.0}
  \plotone{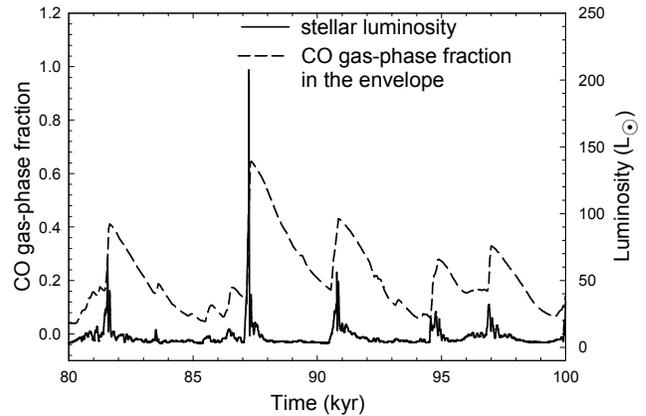}
 \caption{\small Predicted CO gas-phase fraction $\xi_{\rm CO}$ (dashed lines) and total stellar luminosity 
$L_\ast$ (solid lines) vs.~time elapsed since the formation of the protostar.
The correlation between $\xi_{\rm CO}$ and $L_\ast$ is evident.
In particular, $\xi_{\rm CO}$ steeply rises during the burst to a maximum value and 
gradually declines to a minimum value after the burst. The relaxation time to the
pre-burst stage is notably longer than the burst duration. Adapted from {\it Vorobyov et al.}, (2013).\label{fig:cooutburst}}  
 \end{figure}

{\it 3.3.4. Impact on chemistry.}
Large changes in the accretion luminosity of young stars due to episodic 
accretion can drive significant chemical changes in the surrounding core and 
disk.  Several authors have published chemical models exploring these effects 
and have shown that the CO ice evaporates into the gas-phase in the surrounding envelope 
during episodes of increased luminosity, affecting the abundances of many other species through 
chemical reactions (e.g., {\it Lee}, 2007; {\it Visser and Bergin}, 2012; 
{\it Vorobyov et al.}, 2013).  Many of these effects can endure long after an 
accretion burst has subsided, leading to non-equilibrium chemistry compared 
to that expected from the currently observed protostellar 
luminosity. In particular, the abundance of gas-phase CO in the envelope 
can be used as an indicator of past accretion bursts and perhaps 
even a means of measuring the time since the last burst (Fig.~\ref{fig:cooutburst}; {\it Vorobyov 
et al.}, 2013).  Episodic accretion bursts can also affect the abundances and 
chemical compositions of various molecular ices frozen onto dust grains.
{\it Kim et al.}, (2012) showed that a chemical evolutionary 
model including episodic accretion could provide the necessary thermal processing 
 and match the 15.2 $\mu$m CO$_2$ ice absorption features of low-luminosity protostars.

\bigskip
\noindent \textbf{3.4. Disk-Magnetosphere Interactions}
\bigskip

If a protostar has a strong  magnetic field, then the
disk-magnetosphere interaction can lead to episodic accretion, variabilities at
different time-scales, and outflows. In most cases, the magnetic fields of protostars
are not known. However, the measurements of the field in several CTTS indicate the
presence of a magnetic field of a few kG (e.g., {\it Johns-Krull}, 2007; {\it Donati
et. al.}, 2008). The possible presence of a 1~kG magnetic field in 0.05 AU of FU Ori was
 reported by {\it Donati et al.}, (2005) and {\it Green et al.}, (2013a) estimated the
magnetic field of HBC 722 to be 2.2--2.7 kG. There is also
indirect evidence of a magnetic field from the X-ray emissions of FUors and EXors.
If a star has a magnetic field of a few kG, then the disk will be truncated by the magnetosphere at distance $r_m$,
where the magnetic stress  is equal to the matter stress in the
disk ({\it K\"onigl}, 1991).
 In EXors, this radius may be as large as a few stellar radii, while in FUors
 it can be smaller than one stellar radius, or the field may be buried. Numerical simulations predict
 that even in the case of a tiny magnetosphere, the magnetic flux of the star is not buried, but
rather partially
 inflated into the corona, and may drive strong outflows ({\it Lii et al.}, 2012).

{\it 3.4.1. Cyclic accretion in a weak ``propeller" regime.}
 A  model of cyclic accretion was proposed for stars accreting in a
 weak propeller regime. In the propeller regime, the magnetosphere rotates more rapidly than the inner disk, and matter of the disk can be ejected to outflows by a rapidly-rotating magnetosphere  ({\it Illarionov and Sunyaev}, 1975; {\it Lovelace et al.}, 1999). However, in a weak propeller regime, the magnetospheric radius, $r_m$,  is only slightly larger than the corotation radius, $r_c$ (where the angular velocity of the Keplerian disk is equal to the angular velocity of the star), and  such a weak propeller cannot drive outflows. Instead, the star transfers its excess angular momentum to the disk, matter
accumulates in the disk for a long period of time, and a ``dead disk" is formed; then,
part of this disk matter accretes to the star, the magnetosphere expands and the
process repeats in a cyclic fashion ({\it Baan}, 1977; {\it Sunyaev and Shakura}, 1977;
{\it Spruit and Taam}, 1993). The time-scale of accretion episodes is determined by the
accretion rate in the disk and other parameters. {\it D'Angelo and Spruit}, (2010,
2012)  investigated this model  for a wide range of parameters at which
accretion is either cyclic or steady and showed that the model can explain EXor
outbursts. 

{\it 3.4.2. Outflows from the disk-magnetosphere boundary.}
Different models have been proposed to explain high-velocity winds of FUors and EXors (see Sect.~4.4), including: the disk winds
model, where matter is driven by centrifugal force along the inclined field lines of
the disk (e.g., {\it Blandford and Payne}, 1982; {\it Zanni et al.}, 2007);
accretion-powered stellar winds (e.g., {\it Matt and Pudritz}, 2005); and the X-winds,
which are launched centrifugally from the disk-magnetosphere boundary ({\it Shu et
al.}, 1994). The reader is encouraged to read the accompanying chapter by {\it Frank et al.}.

\textit{Conical Winds.} Recent numerical simulations show a new type of wind which
can be important in the cases of EXors and FUors. These winds form at the
disk-magnetosphere boundary during the episodes of high accretion rate. The newly-incoming matter compresses the magnetosphere of the star, the
field lines inflate due to differential rotation between the disk and the star, and
conically-shaped winds flow from the inner disk ({\it Romanova et al.}, 2009; {\em
Kurosawa and Romanova}, 2012). These winds are driven  by the magnetic force, $F_M\sim
- \nabla (rB_\phi)^2$, which arises due to the wrapping of the field lines above the
disk ({\it Lovelace et al.}, 1991). They are also gradually collimated by the magnetic
hoop-stress, and can be strongly collimated for high accretion rates ({\em
Lii et al.}, 2012).  Moreover, the star can
rotate much more slowly than the inner disk (at $r_m \ll r_c$).
This is different from the X-winds, which
require the condition of $r_m\approx r_c$. Conical winds appear during a burst of
accretion and continue for the entire duration of the burst. A magnetic field of a few
kG is required for FUors, while in EXors the field can be weaker.
The conical wind model was compared with the empirical
model based on the spectral analysis of the winds in FU Ori ({\it Calvet et al.}, 1993;
{\it Hartmann and Calvet}, 1995). A reasonably good agreement was achieved between
these models ({\it K\"onigl et al.}, 2011).

\textit{Propeller-driven Winds.} If a protostar rotates much more rapidly than the
inner disk (strong propeller regime)
 and the accretion rate in the disk is relatively high,
then a significant part of the disk can be redirected to the outflows by the
rapidly-rotating magnetosphere
 ({\it Romanova et al.}, 2005, 2009; {\it Ustyugova et al.}, 2006). In this regime, accretion and outflows also
occur in cycles, where matter accumulates in the inner disk, diffuses through the
field lines of the rapidly-rotating magnetosphere, and is ejected to the outflows; then, the magnetosphere expands and
the cycle repeats {({\it Goodson et al.}, 1997; {\it Lii et al.}, 2013)}. The
time-scale of the cycle varies from a few weeks to a few months, and depends on a
number of parameters, such as the accretion rate in the disk and the diffusivity at
the disk-magnetosphere boundary.

In the case of a rapidly-rotating star, outflows have a second component: a
magnetically-dominated, low-density and high-velocity jet, where matter is accelerated
rapidly by the magnetic force that appears due to the winding of the stellar field
lines (a ``magnetic tower"). The jet carries significant energy and angular momentum
from the star to the corona, causing the star to spin down rapidly (e.g., {\it
Romanova et al.}, 2005). If protostars accrete most of the mass during episodes of enhanced accretion, then
powerful outflows observed can be associated with the episodes of
strongest accretion (e.g., {\it Reipurth}, 1989). These outflows carry matter and
angular momentum into the cloud and may influence the overall dynamics of
star formation.

{\it 3.4.3. Variability during bursts of accretion.}
During a burst of accretion, different processes are expected to occur at the
disk-magnetosphere boundary. Matter may flow to the star above the magnetosphere in
two ordered funnel streams and form two hot spots on the stellar surface ({\it Bertout
et al.}, 1988; {\it Romanova et al.}, 2004). Alternatively, matter may accrete through
the magnetic Rayleigh-Taylor instability, where several unstable ``tongues" penetrate
the field lines and form spots of chaotic shape and position ({\it Kulkarni and Romanova},
2008; {\it Romanova et al.}, 2008). In these cases, the light curve and spectral changes appear
chaotic, with a few accretion events per period of the inner disk ({\it Kurosawa and
Romanova}, 2013). Observations of young stars often show variability at this time scale
(e.g., {\it Alencar et al.}, 2010; {\it Cody et al.}, 2013). 
In the unstable regime, the  tongues  rotate  with the angular velocity of the inner disk, and  the frequency of the inner disk may be present  in the Fourier spectrum  of the light-curves. Variations of the accretion rate will lead to variations of the inner disk radius and this frequency will vary in time.
In cases of small magnetospheres, $r_m\lesssim (1-2) R_\star$, one or two regular tongues  rotate orderly with the frequency of the inner disk  ({\it Romanova and Kulkarni}, 2009; {\it Bachetti et al.}, 2010). 
This phenomenon can
potentially explain the quasi-periodic variability of 2-9 days, which has been
recently observed in FU Ori and Z CMa by {\it Siwak et al.,} (2013). Alternatively, it
can be connected with the waves excited in the disk by the tilted dipole (e.g., {\it
Bouvier et al.}, 1999; {\it Romanova et al.}, 2013). The modulation of the blueshifted
spectral lines in FU Ori with a period of 14 days ({\it Powell et al.}, 2012) may be a sign
of modulation of the wind by the waves in the disk. Different longer-period
variabilities  (e.g., {\it Hillenbrand et al.}, 2012)
can also be connected with the waves in the disk, excited at different distances from
the star.

Another type of variability may be connected with \textit{episodic inflation} and
reconnection of the field lines connecting a star with the inner disk ({\it Aly and
Kuijpers}, 1990). The signs of such variability on a time-scale of a few stellar
rotations have been observed by {\it Bouvier et al.}, (2003) in  AA Tau and in
young bursting stars  by {\it Findeisen et al.}, (2013).
This process may also lead to the phenomenon of episodic X-ray flares during the
burst.

\bigskip
\centerline{\textbf{4. CURRENT VIEW OF EPISODIC ACCRETION}}
\bigskip

Despite great progress both observationally and theoretically, there is still a great deal of uncertainty concerning the origin of the outbursts,  whether the mechanism is different for FUors and EXors, and whether both classes of outbursting sources can be various examples of the same phenomenon in the time history of a forming star. In this section, we aim to highlight specific characteristics --- such as outburst timescales, object evolutionary stages as inferred from infrared/submm spectral features and SEDs, the presence/absence of outflows, accretion rates, binarity, and high-energy activity levels --- in which classical FUors and EXors, along with analogous newly discovered eruptive objects, converge as well as diverge. However, we caution that  many critical behaviors are not easily disentangled, and the quest to understand the origin of pre-main-sequence outbursts will still require sustained efforts in the coming years.

\bigskip
\noindent
\textbf{4.1.
Outburst Timescales and Repetition}
\bigskip

The recently discovered examples of outbursts have begun to fill in the notional gap between long-duration (classical FUor) and short-duration (classical EXor) eruptions (Fig.~\ref{fig:ilight}). Several objects (e.g., OO Ser, V1647 Ori) appear to display outburst decay times of a few years, i.e., intermediate between the two classes. Such a conceptual progression from a bimodal distribution to a continuum of outburst timescales is hence perhaps merely the natural consequence of the discovery of additional outbursting YSOs and pre-main-sequence stars, combined with the longer baseline and more comprehensive arsenal of observational data available to measure outburst decay times for both previously and newly identified eruptive objects.

On the other hand, the question of a clear distinction between FUor and EXor classes in terms of outburst repetition and duty cycle remains open. The repetitive nature of EXor outbursts  might be considered a defining characteristic of such objects --- one that potentially links strong, short-timescale EXor eruptions to the lesser variability of TTS more generally. In contrast, we simply have not had time to observe the cessation of any of the classical FUor outbursts and, hence, we are unable to assess the potential for repetition of these long-duration eruptions. But essentially all theoretical models predict that FUor outbursts should also occur more than once during early stellar evolution, with an average time span  of thousands of years between outbursts (Sect.~3.2; Fig.~\ref{fig:bursts1}); {\it Scholz et al.}, (2013) estimated time intervals between outbursts of 5--50 kyr.

FUor and EXor systems are often surrounded by ring- or cometary-shaped nebulae that evidently reside in the outflow and/or the parent cloud (e.g., {\it Goodrich}, 1987).
These bright optical/infrared reflection nebulae become illuminated by the central stars as their activity  levels increase (sometimes revealing light-travel-time effects that facilitate distance and luminosity determinations; e.g., {\it Brice\~no et al.}, 2004). Furthermore, HH objects and jets tend to be associated with sources that have recently entered elevated activity states (e.g., {\it Reipurth and Aspin}, 1997; {\it Takami et al.}, 2006). Thus, outflow structures might be used to infer protostellar and pre-main-sequence outburst duty cycles. 
However, any direct connection between outflow and outburst activity is difficult to infer; e.g., the time intervals between the appearance of large working surfaces in HH flows, typically 500$-$1000 years, imply that HH driving sources are more likely to be outside of a higher-accretion FUor or EXor state ({\it Reipurth and Bally}, 2001). 

\bigskip
\noindent
\textbf{4.2.
Evolutionary Stages}
\bigskip

Just as the outburst duration gap between FUors and EXors has closed, the long-held view that longer duration eruptions involve higher protostellar accretion rates and occur at earlier stages of pre-main-sequence stellar evolution (e.g., {\it Hartmann and Kenyon}, 1996; {\it Sandell and Weintraub}, 2001) has become subject to question. This increasing ambiguity is in large part the result of the recent scrutiny of FUors via higher-resolution infrared and submm imaging and spectroscopy. It is now apparent that classical FUors present a highly heterogeneous set of near- to mid-infrared spectral features and SEDs that could possibly relate to an evolutionary sequence within the FUor class (e.g.,  {\it Quanz et al.}, 2007c). In this scheme, the presence or (partial) depletion of an envelope may significantly modify the observational properties (e.g., silicate in absorption or emission, strong far-infrared excess). However, new outbursting sources have demonstrated that the case may not be as simple: there are often more deeply embedded protostars lying in close proximity to optically-identified  eruptive objects that show FUor characteristics (e.g., {\it Green et  al.,} 2013b; {\it Dunham et al.}, 2012). While some FUors appear relatively devoid of significant circumstellar envelopes, it also appears that some FUors do possess massive, molecule-rich circumstellar envelopes (e.g., {\it K\'osp\'al et al.}, 2011a). The behavior and strength of outbursts may also be influenced by the presence of a wind/outflow and its interaction with a surrounding envelope ({\it Clarke et al.}, 2005). 

Theoretically, it is difficult to provide a single evolutionary framework to explain both EXors and FUors.  The difficulty with providing such a unified scenario is probably due to different causes. First, it is extremely difficult to self-consistently model the evolution of the entire disk, starting from sub-AU to 
hundred-AU scales, and as a result most theoretical models  focus on either the inner 
or outer disk. 
Second, only few attempts have been made up to now to directly compare 
theoretical models to actual observations, e.g., in terms of detailed 
light curve or color evolution modeling for specific events. As the models 
become more sophisticated and observations richer, this is certainly a direction 
to pursue in the future. Third, very little has been done from the theoretical 
standpoint in order to describe EXors. The models of {\it D'Angelo and Spruit}, (2010, 2012) appear to be promising but still lack a detailed time-dependent 
calculation. The explanation of the  origin of EXor outbursts as related to the innermost parts of the disk 
might establish them as a separate class with respect to FUors. 

When focusing on individual theoretical models, those presented in {\it 
Nayakshin and Lodato}, (2012; presented in Sect.~3.2.1) do show variability on a variety of timescales and 
amplitude that might reproduce, within a single scenario, both EXors and FUors. 
The models of {\it Vorobyov and Basu}, (2005, 2006, 2010; presented in Sect.~3.2.3) produce luminosity outbursts 
with amplitudes typical for both FUors and EXors but fail to reproduce the short 
rising times of EXors, possibly due to limitations of the numerical code. 
Certainly, theoretical models do require significant development 
before a proper comparison with data can be made. 

Finally, radiative transfer models should attempt to model the behavior of disk interiors and their atmospheres during outbursts, e.g., as a function of mass accretion rate or disk and envelope masses. Such models should determine whether such parameters can explain the CO bandhead in absorption in FUors but in emission in EXors, and pursue whether such CO features can change depending on the conditions, possibly explaining, e.g., the reversal of CO observed early in the outburst of V1647 Ori.

\bigskip
\noindent
\textbf{4.3.
Binarity}
\bigskip

The search for tight binaries in FUors and EXors is linked with the quest to identify the origin of erupting events. {\it Bonnell and Bastien}, (1992) proposed that FUor outbursts may be due to a perturbation induced by a companion at periastron passage. Noting that there is already a list of known FUor binaries,  {\it Reipurth and Aspin}, (2004b) proposed that FUors may be newborn binaries that have become bound when a small nonhierarchical multiple system breaks up and the two components spiral in toward each other, perturbing their disks.  This model derives particular motivation from the observation that FU Ori is the northern component in a close ($0\farcs 5$), pre-main-sequence binary system  ({\it Reipurth and Aspin}, 2004b; {\it Wang et al.}, 2004) whose non-outbursting component, FU Ori S, is likely more massive than the FUor namesake ({\it Beck and Aspin}, 2012; {\it Pueyo et al.}, 2012).  The scenario advanced by {\it Reipurth and Aspin}, (2004b) implies that FU Ori must be a close binary ($<$ 10 AU; see also {\it Malbet et al.}, 1998, 2005; {\it Quanz et al.}, 2006) and, if so, the newly discovered companion is the outlying member in a triple system. Z CMa, a $0\farcs 1$ FUor/Herbig Be  binary surrounded by a circumbinary disk ({\it Alonso-Albi et al.}, 2009) --- and with jets emanating from both components ({\it Whelan et al.}, 2010; {\it Benisty et al.}, 2010; {\it Canovas et al.}, 2012) --- is another potential example of a system that has undergone binary-disk interactions, although the observed outbursts in this binary do not always stem from FUor variability ({\it Teodorani et al.}, 1997; {\it van den Ancker et al.}, 2004; {\it Szeifert et al.}, 2010; {\it Hinkley et al.}, 2013). 

If such binary-disk interaction is the dominant mechanism to trigger FUor outbursts, then FUor eruptions should preferentially occur in close binaries, i.e., in about 20\% of all stars. However, V1057 Cyg and V1515 Cyg, both ``classical'' FUors, are not known to harbor close companions, although it is difficult to probe the 1--10 AU separation range, the most relevant for the binarity model, due to their distances. In the case of EXors, some stars are known visual or spectroscopic binaries (e.g., V1118 Ori at 72 AU separation, {\it Reipurth et al.}, 2007b; UZ Tau E with $a\sim 0.15$~AU, {\it Prato et al.}, 2002; EX Lup might also harbor a brown dwarf companion, {\it K\'osp\'al}, priv. comm.), while others show no evidence of binarity (e.g., {\it Melo}, 2003; {\it Herbig}, 2007, 2008). 

In summary, the jury is still out on whether binarity plays any role in triggering eruption events in EXors or FUors;  further studies aiming at discovering faint companions or possibly planets in disks of erupting stars are clearly needed in the coming years. The reader is encouraged to read the accompanying chapter by {\it Reipurth et al.}

\bigskip
\noindent
\textbf{4.4.
Accretion, Infall, Winds and Outflows}
\bigskip

Pre-main sequence accretion rates are  difficult to determine and, for a given class of object, are determined via a variety of means and, thus, highly inhomogeneous. Eruptive stars are no exception to this rule. For instance, the photometric data and flux-calibrated spectra of the outburst phase can be compared with SED models to constrain the accretion disk parameters. Correlations between mass accretion rates and the
emission line fluxes, obtained in the framework of magnetospheric accretion, can also be used.

Nevertheless, some general statements can be made  about accretion rates.
The most luminous FUors have mass accretion rates that can reach $10^{-4}-10^{-3}$~\msunyr\  (see Tab.~\ref{tab:fuorexor}). However, eruptive FUor-like objects with lower luminosities imply lower mass accretion rates, as low as 10$^{-6}$~\msunyr. Hence, there is a clear overlap between the ranges of mass accretion rate observed during outburst in
FUors and classical EXors (10$^{-8}$--10$^{-6}$~\msunyr) or intermediate objects  ($\sim10^{-5}$~\msunyr). Quiescent mass accretion rates can start as low $10^{-10}$~\msunyr\  (e.g., {\it Sipos et al.}, 2009) for EXors with little or no envelope, i.e., probing
 episodic accretion in later evolutionary stages. In summary, the peak mass accretion rate is likely not the only physical parameter that determines the nature  of the eruption.

In optical/infrared spectra,   EXors ubiquitously show inverse P Cyg profiles due to infall (e.g., {\it Herbig}, 2007). In addition, 
blueshifted absorption and/or P-Cyg profiles are observed in  spectral lines such as 
H$\alpha$ and Na I D, indicating strong winds. The wind velocities are typically $-50$ to $-300$~\UNITVEL, with maximum values up to $-600$~\UNITVEL  in FUors (e.g., {\it Croswell et al.}, 1987; {\it Vacca et al.}, 2004; {\it Reipurth and Aspin}, 2004a). The maximum velocities appear generally lower in EXors, although some fast wind can be detected as well (e.g., V1647 Ori: {\it Aspin and Reipurth}, 2009). In addition, blueshifted absorption is stronger in FUors than in EXors (e.g., {\it Herbig}, 2007, 2008, 2009). Many FUors also show millimeter signatures of CO outflows. The typical velocity and mass loss rate of the outflows are 10$-$40 \UNITVEL\ and 10$^{-8}-$10$^{-6}$~\UNITSOLARMASS yr$^{-1}$, respectively ({\it Evans et al.}, 1994). However, some objects do not show significant CO emission associated with outflows. As discussed in Sect. 4.1, the connection between HH objects and outbursts is also difficult to infer. However, the strength of the disk wind increases during EXor outbursts (e.g., EX Lup; {\it Sicilia-Aguilar et al.}, 2012) and decays as the outburst decays (e.g., {\it Aspin et al.}, 2010). This relation indicates that the wind strength is, indeed, related to the mass accretion rate, as generally observed in CTTS ({\it Calvet}, 1997).

\bigskip
\noindent
\textbf{4.5.
High-Energy Processes}
\bigskip

The scant existing high-energy data  make apparent that episodes of high accretion rate, whether of sustained or more transient nature, are usually accompanied by enhanced X-ray emission (see Sect.~2.10). In the case of sustained outbursts (i.e., for ``classical'' FUors and V1647 Ori), the emission (when detected) is overall relatively hard, betraying an origin in magnetic activity. For shorter-duration outbursts (as in the ``classical'' EXors), at least some of the enhanced X-ray flux appears to arise in accretion shocks.  V1647 Ori appears to represent something of a hybrid case, in that its emission during high-accretion states is dominated by plasma that is too hot to be due to accretion shocks, yet is confined to ``hot spots'' very near the stellar surface  ({\it Hamaguchi et al.}, 2012). Further X-ray observations are needed to more firmly establish whether and how the X-ray flux levels and plasma temperatures of eruptive young stars correlate with both long- and short-term variations in optical/infrared fluxes and other (e.g., emission-line-based) accretion and outflow signatures. 

\bigskip
\noindent
\centerline{\textbf{5. FUTURE DIRECTIONS}}
\bigskip

We identify here a number of potentially interesting directions that will or should be explored in the field of episodic accretion:

$\bullet$ Continuum and molecular line images with ALMA will provide new opportunities to firmly establish the envelope vs.\ disk masses of FUors and EXors, so as to compare with each other and with those of deeply embedded protostars and CTTS, and to study the chemistry taking place in disks and being modified due to episodic accretion events.

$\bullet$ The output of present and forthcoming generations of  large-field optical monitoring facilities (e.g., Digitized Sky Survey, Palomar Transient Factory, Large Synoptic Survey Telescope) will continue to  enlarge the sample of eruptive pre-main-sequence objects. We can potentially take advantage of these data  to deduce the frequency of eruptive objects, and determine accretion burst duty cycles, as functions of mass and class. However, these identifications will not include protostars at very early (cloud- and/or envelope-embedded) pre-main-sequence evolutionary stages. Nor will they allow continuous time monitoring to study time variability over long timescales.

$\bullet$ Our present understanding of episodic accretion is potentially heavily biased, due to our ``traditional'' reliance on optically-detected eruptions in identifying FUors and EXors. The identification of eruptions in the near-infrared has begun to mitigate this bias somewhat, mainly thanks to the 2MASS survey. {\it WISE} has now provided an all-sky mid-infrared snapshot against which future wide-field mid-infrared imaging surveys can be compared, so as to identify eruptions associated with much more deeply embedded protostars (see {\it Antoniucci et al.}, 2013;  {\it Johnstone et al.}, 2013; {\it Scholz et al.}, 2013). Armed with such identifications, we can begin to more accurately pinpoint the epoch of onset of episodic accretion during protostellar evolution, and obtain follow-up observations from the ground or in space.

$\bullet$ It is worth considering the extent to which dramatic accretion-driven outbursts effectively cause young stars to ``revert'' to earlier stages of protostellar evolution. In other words, if we observe a  Class II object or flat-spectrum source enter a FUor outburst (e.g., HBC 722), are we in effect seeing a born-again Class I protostar? 

$\bullet$ The understanding of outburst feedback  on the inner disk structure (crystallization, chemistry) would profit from  further investigation, especially in the region of the formation of Earth-like planets. Such outbursts may, indeed, have taken place in the history of our Solar system. High angular resolution observations will, thus, help discern structure and physical conditions in the inner disk and search for very close companions.

$\bullet$ Further modeling of the effect of episodic accretion on the disk structure should be considered. The CO spectrum in absorption observed continuously in FUors but rarely or transiently in EXors has traditionally been explained by heating of the disk interior during the accretion event, assuming built-up material falling from the envelope. However, some FUors, including  FU Ori, do not show evidence of massive envelopes. Thus, it remains unclear why they show these typical FUor spectral characteristics. 

$\bullet$ Finally, it could be worthwhile to investigate numerically the link between EXors and FUors by treating the inner and outer disk simultaneously, although this may be out of reach in the near (and mid) term.

These and other future efforts should continue to focus on the fundamental physical processes underlying outbursts, such as narrowing down the possible mechanisms that can lead to accretion bursts, identifying the key system parameters that control burst energetics (amplitude, duration, repetition), and constraining the range of key system parameters such as accretion rates, outflow rates, and star/disk/outflow geometries. 

\vspace*{5mm}

\textbf{Acknowledgments.}  We dedicate this review to the late George~H.~Herbig, who passed away on October 12, 2013, for his pioneering and long-lasting work on eruptive young stars. We acknowledge the fantastic works of a long list of authors mentioned in the reference list that have contributed to our knowledge of episodic accretion in star and planet formation, among them Bo Reipurth whom we wish to thank for carefully reading  the manuscript and providing detailed comments as referee of this chapter. Henrik Beuther is also thanked for providing further editorial comments to improve the manuscript. 
We  are grateful to Michael Richmond for providing the V1647 Ori light curves in Fig.~\ref{fig:xrayoptirlc} and M\'aria Kun for reading the manuscript and providing corrections. Finally we thank several PPVI participants for coming forward and providing useful comments and feedback to improve this review, among them William Fischer, Chris McKee, and Stella Offner.  

We have attempted to include all refereed publications from 1996 until 2013 that are relevant to the study of episodic accretion. We apologize if we have unintentionally missed publications. MMR was supported by NSF grant AST-1211318,  GL by  PRIN MIUR 2010-2011, project 2010LY5N2T, ZZ by NASA HST-HF-51333.01-A, and P\'A and \'AK  partly by OTKA 101393.

\bigskip

\centerline{\textbf{REFERENCES}}
\bigskip
\parskip=0pt
{\small
\baselineskip=11pt

\refs \'Abrah\'am P., et al.~(2004a) {\it Astron. Astrophys., 419}, L39. 

\refs \'Abrah\'am P., et al.~(2004b) {\it Astron. Astrophys., 428}, 89. 

\refs \'Abrah\'am P., et al.~(2006) {\it Astron. Astrophys.,  449}, L13.

\refs \'Abrah\'am P., et al. (2009) {\it\nat, 459}, 224.

\refs Acosta-Pulido J.~A., et al. (2007) {\it Astron. J. 133}, 2020.

\refs Adams F.~C., et al. (1988) {\it Astrophys. J., 326}, 865.

\refs  Alencar S.~H.~P., et al. (2010) {\it Astron. Astrophys., 519}, A88.

\refs Alonso-Albi T., et al. (2009) {\it Astron. Astrophys.,  497}, 117.

\refs Aly J.~J. and Kuijpers J. (1990) {\it Astron. Astrophys., 227}, 473.

\refs Andrews S.~M. and Williams J.~P.~(2005)  {\it \apj, 631}, 1134.

\refs Andrews S.~M., et al. (2004)  {\it \apj, 610}, L45.

\refs Antoniucci S., et al. (2013) {\it\na, 23,}  98.

\refs Armitage P.~J., et al. (2001) {\it Mon. Not. R. Astron. Soc., 324}, 705.

\refs Aspin C. (2011a) {\it Astron. J., 141}, 196.

\refs {Aspin} C. (2011b) {\it\aj, 142}, 135.

\refs Aspin C. and Reipurth B. (2003) {\it Astron. J., 126}, 2936.

\refs Aspin C. and Reipurth B. (2009) {\it Astron. J., 138}, 1137.

\refs Aspin C. and Sandell G. (1994) {\it Astron. Astrophys., 288}, 803.

\refs Aspin C. and Sandell G. (2001) {\it Mon. Not. R. Astron. Soc., 328}, 751.

\refs {Aspin} C., et al. (2006) {\it\aj, 132,} 1298.

\refs Aspin C., et al. (2008) {\it\aj, 135,} 423.

\refs Aspin C., et al.\ (2009a) {\it\aj, 137,} 431.

\refs Aspin C., et al.~(2009b)  {\it \aj, 137}, 2968.

\refs {Aspin} C., et al. (2009c) {\it\apj, 692,}  L67.

\refs Aspin C., et al.\ (2010) {\it\apj, 719}, L50. 

\refs Audard M., et al. (2005) {\it Astrophys. J., 635}, L81.

\refs Audard M.,  et al. (2010) {\it Astron. Astrophys., 511}, A63.

\refs Baan W. (1977) {\it\apj, 214}, 245.

\refs Bachetti M., et al. (2010) {\it\mnras, 403}, 1193.

\refs Bae J., et al. (2013) {\it Astrophys. J., 764}, 141.

\refs Balbus S.~A. and Hawley J.~F.\ (1998) {\it Rev.  Modern Phys., 70,} 1.

\refs Banzatti A., et al. (2012a) {\it Astrophys. J., 745},  90.

\refs Banzatti A., et al. (2012b) {\it Astrophys. J., 751},  160.

\refs Baraffe I., et al. (2009) {\it Astrophys. J., 702}, L27.

\refs Baraffe I., et al. (2012) {\it Astrophys. J., 756,} 118. 

\refs {Bastien} F.~A., et al. (2011) {\it \aj, 142}, 141.

\refs Basu S. and Vorobyov E. I. (2012) {\it\apj, 750}, 30.

\refs Beck T.~L. and Aspin C.~(2012) {\it \aj, 143}, 55.

\refs Bell K. R. and Lin D. N. C. (1994) {\it Astrophys. J., 427}, 987.

\refs Bell K. R., et al. (1995) {\it Astrophys. J., 444}, 376.

\refs Bell  K.~R., et al. (2000) In {\it Protostars and Planets IV} (V. Mannings et al., eds.), pp.~897-926. Univ. of Arizona, Tucson.

\refs Benisty M., et al. (2010) {\it Astron. Astrophys.,  517}, L3.

\refs Bertin G. and Lodato G. (1999) {\it Astron. Astrophys., 350}, 694.

\refs Bertout C., et al.\ (1988) {\it\apj, 330}, 350. 

\refs Blandford R.~D. and Payne D.~G. (1982) {\it\mnras, 199}, 883.

\refs Boley A.C., et al. (2006) {\it\apj, 651}, 517.

\refs Bonnell I. and Bastien P. (1992) {\it Astrophys. J., 401}, L31.

\refs Bouvier J., et al. (1999) {\it Astron. Astrophys., 349}, 619.

\refs Bouvier J., et al. (2003) {\it\aap, 409,} 169.

\refs {Brice{\~n}o} C., et al.~(2004) {\it\apj, 606,} L123.

\refs Brittain S.~D., et al. (2007) {\it Astrophys. J., 670}, L29.

\refs Brittain S.~D., et al. (2010) {\it\apj, 708,} 109.

\refs Calvet N.\ (1997), In {\it IAU Symp.~182, Herbig-Haro  Flows and the Birth of Stars} (B. Reipurth and C. Bertout, eds.), pp.~417-432. Kluwer,  Dordrecht.

\refs Calvet N., et al. (1991a) {\it Astrophys. J., 383}, 752.
        
\refs Calvet N., et al. (1991b) {\it Astrophys. J., 380}, 617.

\refs Calvet N., et al. (1993) {\it\apj, 402}, 623.

\refs Canovas H., et al. (2012) {\it Astron. Astrophys., 543}, A70.

\refs Caratti o Garatti A., et al. (2011) {\it Astron. Astrophys., 526}, L1.

\refs Caratti o Garatti A., et al. (2013) {\it\aap, 554}, A66.

\refs Casali M.~M.\ (1991) {\it\mnras, 248,} 229.

\refs Cha S.-H. and Nayakshin S. (2011), {\it Mon. Not. R. Astron. Soc., 415}, 3319.

\refs Chavarria C.\ (1981) {\it\aap, 101,} 105.

\refs Chen H., et al.\ (1995) {\it\apj, 445,} 377.

\refs Chen W. P., et al. (2012) {\it Astrophys. J., 751}, 118.

\refs Chochol D., et al. (2006) {\it Contrib. Astron. Obs. Skalnate Pleso, 36}, 149.

\refs Chou M.-Y., et al. (2013) {\it\aj, 145}, 108.

\refs Clarke C. J. and Armitage P. J. (2003) {\it Mon. Not. R. Astron. Soc., 345} 691.

\refs Clarke C.~J. and Syer D.\ (1996) {\it\mnras, 278,} L23.

\refs Clarke C. J., et al. (1990) {\it Mon. Not. R. Astron. Soc., 242}, 439. 

\refs Clarke C. J., et al. (2005) {\it Mon. Not. R. Astron. Soc., 361}, 942.

\refs Cody A.-M., et al. (2013), {\it\aj, 145}, 79.

\refs Coffey D., et al. (2004) {\it\aap, 419}, 593.

\refs Cohen M., et al. (1981) {\it   Astrophys. J., 245}, 920.

\refs Cohen M., et al. (1983) {\it Astrophys. J., 273}, 624.

\refs Connelley M.~S. and Greene T.~P.\ (2010) {\it\aj, 140,} 1214.

\refs Cossins P., et al. (2009) {\it\mnras, 393}, 1157.

\refs Cossins P., et al. (2010) {\it\mnras, 401}, 2587.

\refs Costigan G., et al. (2012) {\it\mnras, 427}, 1344.

\refs Covey K. R., et al. (2011) {\it Astron. J., 141},  40.

\refs Croswell K., et al. (1987) {\it\apj, 312}, 227.

\refs D'Angelo C.~R. and Spruit H.~C. (2010) {\it\mnras, 406}, 1208. 

\refs D'Angelo C.~R. and Spruit H.~C. (2012) {\it\mnras, 420}, 416. 

\refs Dent W.~R.~F., et al.\ (1998) {\it\mnras, 301,} 1049.

\refs Donati J.~F., et al. (2005) {\it\nat, 438}, 466.

\refs Donati J.-F., et al. (2008) {\it\mnras, 386}, 1234.

\refs Dunham M.~M.~and Vorobyov E.~I.~(2012)  {\it\apj, 747}, 52.

\refs Dunham M.~M., et al.~(2010)  {\it\apj, 710}, 470.

\refs Dunham M.~M., et al. (2012) {\it\apj, 755}, 157.

\refs Dunham M.~M., et al.\ (2013) {\it\aj, 145,} 94.

\refs Durisen R.~H., et al. (2007) In {\it Protostars and Planets V} (B. Reipurth, D. Jewitt, and K. Keil, eds.), pp.~607-622. Univ. of Arizona, Tucson.

\refs Eisl{\"o}ffel J., et al.\ (1990) {\it \aap, 232,} 70. 

\refs Eisl{\"o}ffel J., et al. (1991) {\it Astrophys. J., 383}, L19.

\refs Eisner J.~A. and Hillenbrand L. A. (2011) {\it Astrophys. J., 738}, 9.

\refs Elias J.~H.\ (1978) {\it\apj, 224,} 857.

\refs Enoch M.~L., et al.~(2009) {\it \apj, 692}, 973.

\refs Evans N.~J.~II, et al. (1994) {\it Astrophys. J., 424}, 793.

\refs Evans N.~J.~II, et al.~(2009) {\it\apjs, 181}, 321.

\refs Fedele D., et al. (2007) {\it Astron. Astrophys., 472}, 207.

\refs Feigelson E.~D. and Montmerle T.\ (1999) {\it \araa, 37,} 363.

\refs Findeisen K., et al.\ (2013) {\it\apj, 768,} 93.

\refs Fischer W.~J., et al. (2012) {\it Astrophys. J., 756}, 99.

\refs Gammie C. (1996) {\it Astrophys. J., 457}, 355.
         
 \refs Gammie C. (2001)  {\it Astrophys. J., 553}, 174.       

\refs Giannini T., et al. (2009) {\it Astrophys. J., 704}, 606.

\refs Gibb A.~G.\ (2008) In {\it Handbook of Star Forming Regions, Volume I: The Northern Sky} (ed. B. Reipurth), pp.~693-731. ASP Monograph Publications, San Francisco.

\refs {Gibb} E.~L., et al. (2006) {\it\apj, 641,} 383.

\refs Goodrich R. W. (1987) {\it\pasp, 99}, 116.

\refs Goodson A. P., et al. (1997), {\it\apj, 489}, 199.

\refs Goto M., et al. (2011) {\it Astrophys. J., 728},  5.

\refs Graham J.~A. and Frogel J.~A.\ (1985) {\it\apj, 289,} 331.

\refs Gras-Vel\'azques \`A. and Ray T. P. (2005) {\it\aap, 443}, 541.

\refs Green J.~D., et al. (2006) {\it Astrophys. J., 648}, 1099.

\refs Green J.~D., et al. (2011) {\it Astrophys. J., 731}, L25.

\refs Green J.~D., et al. (2013a) {\it Astrophys. J., 764}, 22.

\refs Green J.~D., et al.\ (2013b) {\it\apj, 772,} 117.

\refs Greene T.~P., et al.\ (1994) {\it\apj, 434,} 614. 

\refs Greene T.~P., et al. (2008) {\it Astron. J., 135}, 1421.

\refs Grosso N., et al. (2005) {\it Astron. Astrophys., 438}, 159.

\refs Grosso N., et al. (2010) {\it Astron. Astrophys., 522}, A56.

\refs Haas M., et al.\ (1990) {\it\aap, 230,} L1.

\refs Haisch K.~E. Jr., et al.\ (2004) {\it\aj, 127,} 1747. 

\refs Hamaguchi K., et al. (2010)  {\it Astrophys. J., 714,} L16.

\refs Hamaguchi K., et al. (2012) {\it Astrophys. J., 741}, 32.

\refs Hartigan P. and Kenyon S.~J.\ (2003) {\it\apj, 583,} 334.

\refs Hartmann L. (1991) In {\it Physics of Star Formation and Early Stellar Evolution. NATO Adv. Study Inst.} (C. J. Lada and N. D. Kylafis, eds.), pp.~623-648. Kluwer, Dordrecht.

\refs Hartmann L. (2008) In {\it Accretion Processes in Star Formation}. Cambridge University Press, Cambridge.

\refs Hartmann L. and Calvet N. (1995) {\it\aj, 109}, 1846.

\refs Hartmann L. and Kenyon S. J. (1985) {\it Astrophys. J., 299}, 462.

\refs Hartmann L. and Kenyon S. J. (1987a)  {\it Astrophys. J., 322}, 393.

\refs Hartmann L. and Kenyon S. J. (1987b) {\it Astrophys. J., 312}, 243.    

\refs Hartmann L. and Kenyon S. J. (1996) {\it Ann. Rev. Astron. Astrophys., 34}, 207.

\refs Hartmann L., et al.\ (1989) {\it\apj, 338,} 1001.

\refs Hartmann L.,  et al. (1993) In {\it Protostars and Planets III} (E. H. Levy and J. I. Lunine, eds.), pp.~497--520. Univ. of Arizona, Tucson.

\refs Hartmann L., et al. (2004) {\it Astrophys. J., 609}, 906.

\refs Henning, Th., et al. (1998) {\it\aap\, 336,} 565.

\refs Herbig G.~H. (1966) {\it Vistas Astron., 8}, 109.

\refs Herbig G.~H. (1977) {\it\apj, 217}, 693.

\refs Herbig G.~H. (1989) In {\it ESO Workshop on Low Mass Star Formation and Pre-Main Sequence Objects} (B. Reipurth, ed.), pp.~233-246. ESO, Garching.

\refs Herbig G.~H. (2007) {\it Astron. J., 133}, 2679.

\refs Herbig G.~H. (2008) {\it Astron. J., 135}, 637.

\refs {Herbig} G.~H. (2009) {\it \aj, 138}, 448.

\refs Herbig G. H., et al. (2001) {\it Publ. Astron. Soc. Pac., 133}, 1547.

\refs Herbig G.~H., et al.~(2003) {\it \apj, 595}, 384.

\refs Hillenbrand L. (2009), In {\it IAU Symp.~258, The Age of Stars}, (E. Mamajek, et al., eds.), pp.~81-88. Cambridge University Press, Cambridge.

\refs Hillenbrand L.~A., et al.~(2013) {\it \aj, 145}, 59.

\refs Hinkley, S., et al. (2013) {\it Astrophys. J., 763}, L9.

\refs Hodapp K. W. (1999) {\it Astron. J., 118}, 1338.

\refs Hodapp K. W., et al. (1996) {\it   Astrophys. J., 468}, 861.

\refs {Hodapp} K.~W., et al. (2012) {\it \apj, 744}, 56.

\refs Hosokawa T., et al. (2011) {\it Astrophys. J. 738,} 140.

\refs Illarionov A. F. and Sunyaev R. A. (1975) {\it\aap, 39,} 185.

\refs Jensen E. L., et al. (2007) {\it\aj, 134}, 241.

\refs Johns-Krull C. M. (2007) {\it\apj, 664}, 975.

\refs Johnson B. M. and Gammie C. F. (2003) {\it Astrophys. J., 597}, 131.

\refs Johnstone D., et al. (2013) {\it\apj, 765,} 133.

\refs Juh\'asz A., et al. (2012) {\it Astrophys. J., 744}, 118.

\refs Kastner J.~H., et al. (2004) {\it Nature, 430}, 429.

\refs Kastner J.~H., et al. (2006) {\it Astrophys. J., 648}, 43.

\refs Kenyon S.~J. (1995a) {\it Astrophys. Space Sci., 233}, 3.

\refs Kenyon S.~J. (1995b) {\it Rev. Mex. Astron. Astrophys. (Conf. Ser.), 1}, 237.

\refs Kenyon S.~J. and Hartmann L.~W. (1991) {\it Astrophys. J., 383}, 664.

\refs Kenyon S.~J. and Hartmann L.~W. (1995)  {\it \apjs, 101}, 117.

\refs Kenyon S. J., et al. (1988) {\it Astrophys. J., 325}, 231.

\refs Kenyon S.~J., et al.~(1990) {\it \aj, 99}, 869.

\refs Kenyon S. J., et al. (1993) {\it\aj, 105}, 1505.

\refs Kenyon S.~J., et al.~(1994) {\it \aj, 108}, 251.

\refs Kenyon S.~J., et al. (2000) {\it Astrophys. J., 531}, 1028.

\refs Kim H.~J., et al.~(2012), {\it Astrophys.~J., 758}, 38. 

\refs Kley W. and Lin D. N. C. (1999) {\it Astrophys. J., 518}, 833.

\refs K\"ohler R., et al. (2006) {\it Astron. Astrophys., 458}, 461.

\refs K\"onigl A.~(1991) {\it\apj, 370}, L39.

\refs K\"onigl A., et al. (2011) {\it\mnras, 416}, 757.

\refs Koresko C. D., et al. (1991) {\it\aj, 102}, 2073.

\refs K\'osp\'al \'A. (2011) {\it Astron. Astrophys., 535}, 125.

\refs K\'osp\'al \'A., et al. (2007) {\it Astron. Astrophys.,  470}, 211.

\refs K\'osp\'al \'A., et al.~(2008) {\it Mon. Not. R. Astron. Soc., 383}, 1015.

\refs K\'osp\'al \'A., et al. (2011a) {\it Astron. Astrophys., 527}, 133.

\refs K\'osp\'al \'A., et al. (2011b) {\it Astrophys. J., 736}, 72.

\refs K\'osp\'al \'A., et al. (2012) {\it Astrophys. J. Supp., 201}, 11.

\refs K\'osp\'al \'A., et al. (2013) {\it   Astron. Astrophys., 551}, 62.

\refs Kratter K. M., et al. (2008) {\it Astrophys. J., 681}, 375.

\refs Kravtsova A. S., et al. (2007) {\it   Astron. Letters 33}, 755.

\refs Kryukova E., et al.~(2012) {\it \aj, 144}, 31.

\refs Kulkarni A. K. and  Romanova M. M. (2008) {\it\apj, 386,} 673.

\refs Kun M., et al. (2011a) {\it Astrophys. J., 733},  L8.

\refs Kun M., et al. (2011b) {\it   Mon. Not. R. Astron. Soc., 413}, 2689.

\refs Kurosawa R. and Romanova M. M. (2012) {\it\mnras, 426}, 2901.

\refs Kurosawa R. and Romanova M. M. (2013) {\it\mnras, 431}, 2673.

\refs Kurucz R.~L.,  et al.  (1974).  Smithsonian Institution, Washington.

\refs Lee J.-E. (2007) {\it J. Korean Astron. Soc., 40}, 83.

\refs Lehmann T., et al. (1995), {\it  Astron. Astrophys., 300}, L9.

\refs Leinert Ch. and Haas M. (1987) {\it Astron. Astrophys., 182}, L47.

\refs Lii P., et al. (2012) {\it\mnras, 420}, 2020.

\refs Lii P.~S, et al. (2013) {\it\mnras}, arXiv:1304.2703.

\refs Lin D. N. C., et al. (1985) {\it Mon. Not. R. Astron. Soc, 212}, 105.

\refs Lodato G. and Bertin G. (2001) {\it Astron. Astrophys., 375}, 455.

\refs Lodato G. and Bertin G. (2003) {\it Astron. Astrophys., 408}, 1015.

\refs Lodato G. and Clarke C. J. (2004) {\it Mon. Not. R. Astron. Soc., 353}, 841.

\refs Lodato G. and Rice W.~K.~M. (2004) {\it Mon. Not. R. Astron. Soc., 351}, 630.

\refs Lodato G. and Rice W.~K.~M. (2005) {\it Mon. Not. R. Astron. Soc., 358}, 1489.

\refs Lodato G., et al. (2007) {\it Mon. Not. R. Astron. Soc., 374}, 590.

\refs Lombardi M., et al. (2008) {\it Astron. Astrophys., 480}, 785.

\refs Lorenzetti D., et al. (2000), {\it Astron. Astrophys., 357}, 1035.

\refs Lorenzetti D., et al. (2006) {\it Astron. Astrophys., 453}, 579.

\refs Lorenzetti D., et al. (2007)  {\it Astrophys. J., 665}, 1182.

\refs Lorenzetti D., et al.~(2009) {\it \apj, 693}, 1056.

\refs Lorenzetti D., et al. (2011) {\it Astrophys. J., 732}, 69.

\refs Lorenzetti D., et al. (2012)  {\it Astrophys. J., 749}, 188.

\refs Lovelace R. V. E., et al. (1991) {\it\apj, 379}, 696.

\refs Lovelace R. V. E., et al. (1999) {\it\apj, 514}, 368.

\refs Machida M. N., et al. (2011) {\it Astrophys. J., 729}, 42.

\refs Magakian T.~Y., et al.\ (2013) {\it \mnras, 432,} 2685.

\refs Magakian T.~Y., et al.\ (2010) {\it\aj, 139,} 969.

\refs Malbet F., et al. (1998)  {\it Astrophys. J., 507}, L149.

\refs {Malbet} F., et al. (2005) {\it \aap, 437,} 627.

\refs Martin R. G. and Lubow S.~H. (2011)  {\it Astrophys. J., 740}, L6.

\refs Martin  R.~G. and Lubow S.~H.\ (2013) {\it\mnras, 432,} 1616.

\refs Martin R. G., et al. (2012a) {\it Mon. Not. R. Astron. Soc., 420}, 3139.

\refs Martin R. G., et al. (2012b) {\it Mon. Not. R. Astron. Soc., 423}, 2718.

\refs Matt S. and Pudritz R. (2005) {\it\apj, 632}, L135.

\refs Mayer L., et al. (2005) {\it\mnras 363}, 641.

\refs McGehee P. M., et al. (2004) {\it Astrophys. J., 616}, 1058.

\refs McMuldroch S., et al. (1993) {\it Astron. J., 106},  2477.

\refs McMuldroch S., et al. (1995) {\it Astron. J., 110}, 354.

\refs Mej\'{\i}a A. C., et al. (2005), {\it \apj, 619}, 1098.

\refs Melo C.~H.~F.\ (2003), {\it \aap, 410}, 269.

\refs Menten K. M., et al. (2007) {\it Astron. Astrophys., 474}, 515.

\refs Millan-Gabet R., et al.~(2006) {\it \apj, 641}, 547.

\refs Miller A. A., et al. (2011), {\it Astrophys. J., 730}, 80.

\refs Moriarty-Schieven G. H., et al. (2008) {\it Astron. J., 136}, 1658.

\refs Mosoni L., et al. (2013) {\it Astron. Astrophys., 552}, A62.

\refs Movsessian T. A., et al. (2003) {\it Astron. Astrophys., 412}, 147.

\refs Movsessian T. A., et al. (2006) {\it Astron. Astrophys., 455}, 1001.

\refs Muzerolle J., et al. (2005) {\it Astrophys. J., 620}, L107.

\refs Nayakshin S. and Lodato G. (2012) {\it Mon. Not. R. Astron. Soc., 426}, 70.

\refs Offner S.~S.~R. and McKee C.~F.~(2011), {\it Astrophys.~J., 736}, 53.

\refs Offner S. S. R., et al. (2009) {\it Astrophys. J., 703}, 131.

\refs {Ojha} D., et al. (2005) {\it Publ. Astron. Soc. Jap., 57,} 203.

\refs {Ojha} D.~K., et al. (2006) {\it\mnras, 368,} 825.

\refs Okuda T., et al. (1997) {\it Publ. Astron. Soc. Jap., 49,} 679.

\refs Osterloh M. and Beckwith S. V. W. (1995) {\it Astrophys. J. 439}, 288.

\refs Parsamian E. S. and Mujica R. (2004) {\it Astrophysics, 47}, 433.

\refs Parsamian E. S., et al. (2002) {\it Astrophysics, 45}, 393.

\refs Peneva S. P., et al.. (2010) {\it  Astron. Astrophys. 515}, 24.

\refs P\'erez L. M., et al. (2010) {\it Astrophys. J., 724}, 493.

\refs Persi P., et al.\ (2007) {\it \aj, 133,} 1690.

\refs Petrov P.~P. and Herbig G.~H.\ (1992) {\it \apj, 392,} 209.

\refs Petrov P. P. and Herbig G.~H. (2008) {\it Astron. J., 136}, 676.

\refs Petrov P., et al. (1998) {\it Astron. Astrophys., 331}, L53.

\refs Pfalzner S. (2008) {\it Astron. Astrophys., 492}, 735.

\refs Pfalzner S., et al. (2008) {\it Astron. Astrophys., 487}, L45.

\refs Polomski E. F., et al. (2005) {\it Astron. J., 129}, 1035.

\refs Popham R.\ (1996) {\it\apj, 467,} 749. 

\refs Popham R., et al.\ (1996) {\it\apj, 473,} 422.

\refs Powell S. L., et al. (2012), {\it Mon. Not. R. Astron. Soc., 426}, 3315.

\refs Prato L., et al.\ (2002) {\it\apj, 579}, L99.

\refs Preibisch T., et al. (2005) {\it Astrophys. J. Supp., 160}, 401.

\refs Prusti T., et al. (1993)  {\it Astron. Astrophys., 279}, 163.

\refs Pueyo L., et al. (2012) {\it Astrophys. J., 757}, 57.

\refs Quanz S.~P., et al. (2006), {\it Astrophys. J., 648}, 472.

\refs Quanz S.~P., et al. (2007a) {\it Astrophys. J., 656}, 287.

\refs Quanz S.~P., et al. (2007b) {\it Astrophys. J., 658}, 487. 

\refs Quanz S. P., et al. (2007c) {\it Astrophys. J., 668}, 359.

\refs Reipurth B.\ (1989) {\it\nat,  340,} 42.

\refs Reipurth B. (1990) In  {\it IAU Symp.~137, Flare stars in star clusters, associations and the solar vicinity} (L.~V. Mirzoyan, et al., eds.), pp.~229-251. Kluwer, Dordrecht.

\refs Reipurth B. and Aspin C. (1997) {\it Astron. J., 114}, 2700.

\refs Reipurth B. and Aspin C.~(2004a)  {\it \apj, 606}, L119.

\refs Reipurth B. and Aspin C. (2004b) {\it \apj, 608}, L65.

\refs Reipurth B. and Aspin C. (2010) In {\it Victor Ambartsumian Centennial Volume, Evolution of Cosmic Objects through their Physical
Activity}, (H. Harutyunyan, A. Mickaelian, and Y. Terzian, eds.), pp.~19-38. Gitutyun Publishing House, Yerevan.

\refs Reipurth B. and Bally J.\ (2001) {\it\araa, 39}, 403. 

\refs Reipurth B., et al.~(2002) {\it \aj, 124}, 2194.

\refs Reipurth B., et al. (2007a) {\it Astron. J., 133}, 1000.

\refs Reipurth B., et al.\ (2007b) {\it \aj, 134}, 2272.

\refs Reipurth B., et al. (2012) {\it Astrophys. J., 748}, L5.

\refs Rettig T.~W., et al. (2005) {\it Astrophys. J., 626}, 245.

\refs Rice W.~K.~M, et al. (2005) {\it\mnras, 364}, L56.

\refs Rice W.~K.~M., et al. (2011) {\it\mnras, 418}, 1356.

\refs Romanova M.M. and Kulkarni A.K. (2009) {\it\mnras, 398,} 1105.

\refs Romanova M.~M., et al. (2004), {\it\apj, 610}, 920.

\refs Romanova M.~M., et al. (2005) {\it\apj, 635}, L165.

\refs Romanova M.~M., et al.  (2008) {\it\apj, 673}, L171.

\refs Romanova M.~M., et al. (2009) {\it\mnras, 399}, 1802.

\refs Romanova M.~M., et al. (2013) {\it\mnras, 430}, 699.

\refs Sandell G. and Aspin C. (1998) {\it Astron. Astrophys., 333}, 1016.

\refs Sandell G. and Weintraub D. A. (2001) {\it  Astrophys. J. Supp., 134}, 115.

\refs Scholz A., et al. (2013) {\it\mnras, 430}, 2910.

\refs Sch\"utz O., et al. (2005) {\it  Astron. Astrophys., 431}, 165.

\refs Semkov, E. H. and Peneva, S. P. (2012) {\it Astron. Sp. Sci., 338}, 95.

\refs Semkov E.~H., et al. (2010) {\it Astron. Astrophys., 523,} L3.

\refs Semkov E.~H., et al. (2011) {\it Bulg. Astron. J., 15}, 49.

\refs Semkov E.~H, et al. (2012) {\it Astron. Astrophys., 542}, 43.

\refs Semkov, E.~H., et al.\ (2013) {\it \aap, 556,} A60.

\refs Shevchenko V.~S.,et al.\ (1991) {\it \sovast, 35,} 135. 

\refs {Shevchenko} V.~S.,  et al. (1997) {\it\aaps, 124,} 33.

\refs Shu F., et al. (1994) {\it\apj, 429}, 781.

\refs Sicilia-Aguilar A., et al. (2008) {\it Astrophys. J., 673}, 382.

\refs Sicilia-Aguilar A., et al. (2012) {\it Astron. Astrophys. 544}, 93.

\refs Sipos N., et al. (2009) {\it Astron. Astrophys., 507}, 881. 

\refs Siwak  M., et al. (2013) {\it\mnras, 432}, 194.

\refs Skinner S.~L., et al. (2006) {\it Astrophys. J., 643}, 995.

\refs Skinner S.~L., et al. (2009) {\it Astrophys. J., 696}, 766.

\refs Skinner S.~L., et al.  (2010) {\it Astrophys. J., 722}, 1654.

\refs Spruit H.~C. and Taam R.~E. (1993) {\it\apj, 402}, 593.

\refs Stamatellos D. and Whitworth A.~P. (2009a) {\it Mon. Not. R. Astron. Soc., 392}, 413.

\refs Stamatellos D. and Whitworth A.~P. (2009b) {\it Mon. Not. R. Astron. Soc., 400}, 1563.

\refs Stamatellos D., et al. (2011) {\it Astrophys. J., 730}, 32.

\refs Stamatellos D., et al. (2012) {\it Mon. Not. R. Astron. Soc., 427}, 1182.

\refs Staude H.~J. and Neckel Th.\ (1991) {\it\aap, 244,} L13. 

\refs Staude H.~J. and Neckel Th. (1992) {\it Astrophys. J., 400}, 556.

\refs Stecklum B., et al. (2007) {\it Astron. Astrophys., 463}, 621.

\refs Stelzer B., et al. (2009) {\it Astron. Astrophys., 499}, 529.

\refs Strom K. M. and Strom S. E. (1993) {\it Astrophys. J., 412}, L63.

\refs Sunyaev R. A. and Shakura N. I. (1977) {\it Soviet Astron. Lett., 3}, 138.

\refs Szeifert T., et al. (2010) {\it Astron. Astrophys., 509}, L7.

\refs Takami M.,  et al. (2006) {\it\apj, 641}, 357. 

\refs Tapia M., et al. (2006) {\it \mnras, 367,} 513. 

\refs Tassis K. and Mouschovias T.~C.\ (2005) {\it \apj, 618,} 783. 

\refs Teets W. K., et al. (2011) {\it Astrophys. J., 741}, 83.

\refs Teets W. K., et al. (2012) {\it Astrophys. J., 760}, 89.

\refs {Teodorani} M., et al.  (1997) {\it \aaps, 126}, 91.

\refs Toomre A. (1964) {\apj, 139,} 1217.

\refs {Tsukagoshi} T., et al. (2005) {\it Publ. Astron. Soc. Jap., 57}, L21.

\refs Turner N. J. J., et al. (1997) {\it Astrophys. J., 480}, 754.

\refs Umebayashi T.\ (1983) {\it Prog. Theor. Phys., 69,} 480.

\refs Ustyugova G.~V., et al. (2006) {\it\apj, 646}, 304.

\refs Vacca W. D., et al. (2004) {\it Astrophys. J.,  609}, L29.

\refs van den Ancker M. E., et al. (2004) {\it Mon. Not. R. Astron. Soc., 349}, 1516.

\refs Venkata Raman V., et al. (2013) {\it Research in Astron. Astrophys., 13,} 1107.  

\refs Vig S., et al. (2006) {\it Astron. Astrophys., 446}, 1021.

\refs Visser R. and Bergin E. A. (2012) {\it Astrophys. J., 754}, 18.

\refs Vorobyov E.~I. (2009) {\it Astrophys. J., 704}, 715.

\refs Vorobyov E.~I. (2010) {\it Astrophys. J., 729}, 146.

\refs Vorobyov E.~I. (2013) {\it Astron. Astrophys., 552}, 129.

\refs Vorobyov E.~I. and Basu S. (2005) {\it Astrophys. J., 633}, L137.

\refs Vorobyov E.~I. and Basu S. (2006) {\it Astrophys. J., 650}, 956.

\refs Vorobyov E.~I. and Basu S. (2010) {\it Astrophys. J., 719}, 1896.

\refs Vorobyov E.~I., et al.~(2013) {\it Astron.~Astrophys., 557,} A35.

\refs {Walter} F.~M., et al.  (2004) {\it \aj, 128}, 1872.

\refs Wang H., et al. (2004) {\it Astrophys. J., 601}, L83.

\refs Welin G.\ (1976) {\it\aap, 49,} 145.

\refs Whelan E. T., et al. (2010) {\it Astrophys. J., 720}, L119.

\refs Xiao L., et al. (2010) {\it \aj, 139}, 1527.

\refs Young C.~H. and Evans N.~J.~II (2005)  {\it \apj, 627}, 293.

\refs Zanni C., et al..~(2007) {\it Astron. Astrophys., 469}, 811-828

\refs Zhu Z., et al. (2007) {\it Astrophys. J., 669}, 483.

\refs Zhu Z., et al. (2008) {\it Astrophys. J., 684}, 1281.

\refs Zhu Z., et al. (2009a) {\it \apj, 694}, L64.

\refs Zhu Z., et al. (2009b) {\it Astrophys. J., 694}, 1045.

\refs Zhu Z., et al. (2009c) {\it Astrophys. J., 701}, 620.

\refs Zhu Z., et al. (2010a) {\it Astrophys. J., 713}, 1134.

\refs Zhu Z., et al. (2010b) {\it Astrophys. J., 713}, 1143.
}

\end{document}